\documentclass{aastex62}

\usepackage{braket}
\usepackage{bm}
\usepackage{amsmath,amssymb}
\usepackage{subfigure}
\usepackage{xcolor}
\usepackage[normalem]{ulem}
\usepackage{hyperref}

\newcommand{\mode}[2]{\ensuremath{_{#1}\mathrm{S}_{#2}}\,}
\newcommand{\omref}{\ensuremath{\omega_{\rm{ref}}}}
\newcommand{\cL}{\mathcal{L}\,}

\newcommand{\dLB}{\ensuremath{\delta \cL^{B}}\,}
\newcommand{\grad}{\ensuremath{\bm{\nabla}}}

\newcommand{\LamB}{\ensuremath{\Lambda^{B}}}
\newcommand{\bnabla}{\ensuremath{\bm{\nabla}}}
\newcommand{\inner}[2]{\ensuremath{\langle#1|#2\rangle}}
\newcommand{\xiv}{\boldsymbol{\xi}}

\newcommand{\gam}[1]{\ensuremath{\gamma_{#1}}\,}

\newcommand{\wigfull}[6]{\bigg(\begin{smallmatrix} #1 & #2 & #3 \\#4 & #5 & #6 \end{smallmatrix}\bigg)\,}
\newcommand{\wigred}[3]{\bigg(\begin{smallmatrix} \ell' & s & \ell \\ #1 & #2 & #3 \end{smallmatrix}\bigg)\,}
\newcommand{\om}[2]{\ensuremath{\Omega_{#1}^{#2}}\,}
\newcommand{\enc}[1]{\left( #1 \right)\,}
\newcommand{\Bv}{\mathbf{B}\,}
\newcommand{\cH}{\boldsymbol{\mathcal{H}}}
\newcommand{\cG}{\mathcal{G}}
\newcommand{\ev}[1]{\hat{\bm{e}}_{#1}}
\newcommand{\cB}{\mathcal{B}}

\newcommand{\gshvec}[3]{\left(\begin{matrix} #1 \\ #2 \\ #3 \end{matrix}\right)\,}
\newcommand{\cP}{\mathcal{P}^{(\ell)}}

\newcommand{\cK}{\ensuremath{\mathcal{K}}}

\newcommand{\tj}[6]{\left(\begin{array}{ccr}
#1 & #2 & #3\\
#4 & #5 & #6 \end{array}\right)}
\newcommand{\solint}{\int_0^{R_{\odot}} \mathrm{d}r\,}
\newcommand{\solintv}{\int_{\odot} \mathrm{d}^3\mathbf{r}\,}
\newcommand{\cA}{\ensuremath{\mathcal{A}}}

\graphicspath{{./}{figures/}}

\received{\today}
\revised{\today}
\accepted{\today}
\submitjournal{ApJ}

\shorttitle{Frequency splitting due to Lorentz stresses}
\shortauthors{Bharati Das et al.}

\begin{document}

\title{Sensitivity kernels for inferring Lorentz stresses from normal-mode frequency splittings in the Sun}

\correspondingauthor{Srijan Bharati Das}
\email{sbdas@princeton.edu}

\author[0000-0003-0896-7972]{Srijan Bharati Das}
\affiliation{Department of Geosciences \\
Princeton University \\
Princeton, New Jersey, USA}

\author[0000-0002-2365-4723]{Tuneer Chakraborty}
\affil{Department of Astronomy and Astrophysics \\
Tata Institute of Fundamental Research \\
Mumbai, India}

\author[0000-0003-2896-1471]{Shravan M. Hanasoge}
\affil{Department of Astronomy and Astrophysics \\
Tata Institute of Fundamental Research \\
Mumbai, India}

\author[0000-0002-2742-8299]{Jeroen Tromp}
\affiliation{Department of Geosciences \\
and Program in Applied \& Computational Mathematics \\
Princeton University \\
Princeton, New Jersey, USA}



\begin{abstract}
Departures from standard spherically symmetric solar models, in the form of perturbations such as global and local-scale flows and structural asphericities, result in the splitting of eigenfrequencies in the observed spectrum of solar oscillations. Drawing from prevalent ideas in normal-mode coupling theory in geophysical literature, we devise a procedure that enables the computation of sensitivity kernels for general Lorentz stress fields in the Sun. Mode coupling due to any perturbation requires careful consideration of self- and cross-coupling of multiplets. Invoking the isolated-multiplet approximation allows for limiting the treatment to purely self-coupling, requiring significantly less computational resources. We identify the presence of such isolated multiplets under the effect of Lorentz stresses in the Sun.
Currently, solar missions allow precise measurements of self-coupling of multiplets via ``$a$-coefficients" and the cross-spectral correlation signal which enables the estimation of the ``structure coefficients". We demonstrate the forward problem for both self-coupling ($a$-coefficients) and cross-coupling (structure coefficients). In doing so, we plot the self-coupling kernels and estimate $a$-coefficients arising from a combination of deep-toroidal and surface-dipolar axisymmetric fields. We also compute the structure coefficients for an arbitrary general magnetic field (real and solenoidal) and plot the corresponding ``splitting function", a convenient way to visualize the splitting of multiplets under 3D internal perturbations. The results discussed in this paper pave the way to formally pose an inverse problem, and infer solar internal magnetic fields.
\end{abstract}



\section{Introduction} \label{sec:intro}

Solar variability and its impact on space weather phenomena is rooted in the interior dynamics of the Sun \citep{schrijver00}. Large- and small-scale turbulent convective flows \citep{leighton62,Muller92,toomre02solphy,rast03,DRT04} are believed to drive solar dynamo processes \citep{Weiss96,ossendrijver03,miesch05,fan09}, resulting in the evolution of magnetic field. 
Accurate dynamo models have potential to predict solar-cycle variations and, eventually, the planetary-plasma environment \citep{mehsc19,sbdas19}. Today, a plethora of highly sophisticated numerical models attempt to simulate global- and local-scale dynamos \citep{Wicht02,voegler2005,Graham10,hotta15,hotta16}.  Detailed constraints on magnetohydrodynamic (MHD) parameters in the solar interior would help to improve the quality of these numerical models. In this context, helioseismic constraints on the interior magnetic structure of the Sun would prove very useful. 

The MHD Eqns.~\citep[e.g.,][]{goedbloed2004} involve a coupled system of flows and magnetic fields with complicated tensorial terms, making direct seismic inferences mathematically and conceptually challenging. Indeed, observational studies that aim to make sense of local helioseismic measurements in and around magnetic regions have had a controversial history \citep{gizon_etal_2009}. Flows thought to drive the creation and maintenance of active regions have been of significant interest, e.g., determining correlations between flows and magnetic activity has been an active topic \citep{Komm93,Chou2001,BasuAntia03,zhao04}. 

The problem of imaging global magnetic fields through helioseismology has been relatively underexplored, the focus mainly being the solar rotational profile and other global and local flow fields \citep{lavely92,gilesphd,BasuAntia99,zhao04,hanasoge12_conv,hanasoge17_etal}. There have been studies analyzing effects of local fields, such as those contained in a sunspot, on the reduction of wave power \citep{cally00,Schunker06}, and changes in wave speed and flow patterns \citep{gizon_etal_2009,Khomenko15,Svanda14,Rabello-Soares18,Braun19}. Nevertheless, early attempts \citep[e.g.,][]{gough90} that were made to study the impact of global magnetic fields (Lorentz-stress perturbations) on the eigenfrequencies and eigenfunctions of the standard solar model \citep[model S in][]{jcd} focused on the details of the forward problem of frequency shifts induced by an axisymmetric magnetic field not aligned with the rotation axis. \citet{goode04} considered near-surface, small-scale as well as a large-scale deep toroidal fields at the tachocline.
However, lacking a formal relation between Lorentz-stresses and the consequent seismic signatures (sensitivity kernels), these efforts suffered from the drawback of restricting the field geometry to make the problem tractable. Several numerical studies have also considered small perturbations around a magnetized background medium \citep{cameron08,cameron10, hanasoge12_mag,schunker13}.  Attempts at deducing temporal changes in magnetic-field configurations from variations in angular velocity have met with limited success \citep{AntiaChitre13}. In a recent analysis, \cite{cutler} explored the potential of learning about magnetic fields from mode-coupling theory. \cite{hanasoge17} derived analytical forms of the Lorentz-stress sensitivity kernels in the context of normal-mode coupling. This has made it possible to consider the treatment of a completely general magnetic field configuration as a perturbation around a hydrostatic background state. This forms the basis of the current study. 

The eigenmodes of an unperturbed, hydrostatic, spherically symmetric background state are orthogonal and have distinct real eigenfrequencies \citep{goedbloed2004}. Because such a state is azimuthally symmetric, groups of eigenmodes (called multiplets) are degenerate. However,  asymmetric perturbations lift this degeneracy.
For a multiplet isolated in frequency subjected to a zonal perturbation, the split frequency is a function of the azimuthal order~$m$, and can therefore be decomposed in a suitable set of basis polynomials in~$m$ \citep{Sekii1990,Sekii1991,schou_pol_94}.
Frequency splitting due to symmetric flows across the equator \citep{ritzwoller} may be fit using only odd polynomials (these polynomials could be proportional to Legendre polynomials of order~$m$ or Clebsch-Gordon coefficients of the same order). For structure perturbations, such as ellipticity, even-order polynomials span the space of frequency splittings. The choice of this basis polynomial is not unique, but we use the convention followed in \citet{schou_pol_94}. 

The novelty of our work lies in the fact that we present generalized Lorentz-stress sensitivity kernels for frequency splitting. In doing so, we adopt a formalism able to accommodate any choice of field. This is ensured by the technique with which our sensitivity kernels are derived, respecting the second-rank tensorial nature of the Lorentz-stress terms. Sensitivity kernels for similar perturbations are described in terrestrial geophysical literature \citep[e.g.,][]{DT98}. Thus, our kernels allow for systematically inferring Lorentz stresses. Inversions for Lorentz-stresses, just as any other perturbation, are much more precise in near-surface layers than the deep interior. This is because Lorentz-stresses dominate in the near surface as compared to gas pressure, although quickly reversing when moving into deeper layers. Since waves spend most of their time in the near surface, kernels for Lorentz-stress are maximally sensitive to those layers. However, as noted by \cite{gizon_etal_2009}, near-surface localized magnetic features, such as sunspots, induce  pressure, density, and magnetic field perturbations that are too strong to be modelled based on the Born approximation. Therefore the consequent shifts in frequency or wave travel-times are significant. As a result, frequency shifts due to high angular degree magnetic structures may be sensitive to strong near-surface perturbations.

The method of normal-mode coupling to model frequency shifts depends greatly on the proximity of modes in frequency. For closely spaced modes, a strong enough perturbation can cause shifts in frequency which are comparable or larger than the frequency separation between the unperturbed modes. In such a case, one would need to account for cross-coupling of all the modes within a certain frequency window. If, however, the modes are well separated in frequency and perturbations are weak, it may be possible to make do with a self-coupling analysis.
The latter is computationally significantly cheaper and mathematically relatively simple. The modes for which self-coupling is a good approximation are called `isolated'. Groups of such modes, isolated in frequency, that belong to the same degenerate multiplet prior to perturbation are called `isolated multiplets'. In this study we develop a mathematical formalism which respects cross-coupling in the presence of a general Lorentz stress field, and we explore the simplifications associated with self-coupling of isolated multiplets.
We demonstrate the abundance of such isolated multiplets with Lorentz-stress as the only perturbation. We use these sensitivity kernels to estimate the approximate order of magnitude of the frequency splitting expected in the presence of an analytically constructed large-scale Lorentz-stress field. The global strength and configuration of the field is chosen to coarsely resemble realistic models (refer to Section~\ref{sec: Theory3}). Although a more accurate calculation would account for a differentially rotating background state, we do not pursue this here and defer it to a future investigation. Finally, we present a corrected form of the Lorentz-stress sensitivity kernels, which includes terms that were missing in the original work \cite{hanasoge17}. Although those were for mode coupling, the idea is very similar.

In Section~\ref{Sec: Theory1}, we lay out the underlying theoretical tools that are essential for carrying out the perturbation analysis. This includes the background-wave equation and decomposition of the displacement field, magnetic-field vectors and Lorentz-stress tensors in the generalized-spherical-harmonics basis (hereafter GSH).
We formally describe the forward problem in Section~\ref{sec: Theory2} in the context of full-coupling due to a completely general magnetic (or Lorentz-stress) field and thereafter illustrate the method of extracting structure coefficients (and constructing splitting functions) in Section~\ref{sec: coupled_multiplets}. We discuss the simplifications when working with isolated-multiplets in Section~\ref{sec: isolated_multiplet} for axisymmetric as well as non-axisymmetric fields in Section~\ref{Sec: Results}. We first establish the existence of isolated multiplets in Section~\ref{sec: mode_labelling} under the sole influence of Lorentz-stress. Subsequently, we narrow down the problem to self-coupling under axisymmetric fields --- the $a$-coefficient approach.
Kernels for $a$-coefficients are presented in Section~\ref{sec: self-coupling-axisymmetric}. Illustrative calculations for these forward problems using analytically constructed (axisymmetric and non-axisymmetric) magnetic fields are shown in Sections~\ref{sec: Theory3} and \ref{sec: splitting_functions_results}. Discussions and conclusions pertaining to our findings in this study and its comparison to earlier studies of a similar nature can be found in Section~\ref{sec: discussion&conclusion}. 
Appendix~\ref{sec: MHD_linearization} contains the steps leading to the first-order background-wave equation and the Lorentz-stress perturbation operator. The derivation for quasi-degenerate perturbation theory is laid out in Appendix~\ref{sec: Pert_deriv}, followed by a section outlining the analysis of the coupling matrix and the sensitivity kernels in Appendix~\ref{sec: deriv_cpmat}. We present analytical expressions for our custom-designed divergence-free magnetic field in Appendices~\ref{sec:B_construction} and~\ref{sec: H_expr}. We formulate the inverse problem in the GSH basis. The prescription to convert the Lorentz-stress field from the GSH basis to the actual spherical coordinates is stipulated in Appendix~\ref{sec: GSH2SPHR}. Finally, we dedicate Appendix~\ref{sec: a_coeff_Appendix} to $a$-coefficients for the readers' convenience, and Appendix~\ref{sec: DR_mode_labelling} demonstrating the validity of the isolated-multiplet approximation under differential rotation. 

\section{Theoretical formulation}

\subsection{Basic framework and notation} \label{Sec: Theory1}

We begin by considering the full set of coupled ideal MHD equations in CGS units,
\begin{eqnarray}
    \partial_t \rho &=& - \bnabla \cdot (\rho\, \mathbf{v}), \label{eqn: MHD1} \\
    \rho (\partial_t \mathbf{v} + \mathbf{v}\cdot \bnabla \mathbf{v} ) &=& - \bnabla p + c^{-1}\,\mathbf{j}\times \mathbf{B} - \rho \bnabla \phi, \label{eqn: MHD2}\\
    \partial_t p &=& - \mathbf{v}\cdot \bnabla p - \gamma\, p\, \bnabla \cdot \mathbf{v}, \\
    \partial_t \mathbf{B} &=& \bnabla \times (\mathbf{v} \times \mathbf{B}). \label{eqn: MHD4}
\end{eqnarray}
Here $\rho$ denotes the mass density, $\mathbf{v}$ the material velocity, $p$ the pressure, $\phi$ the gravitational potential, $c$ the speed of light,  and $\mathbf{j} = (c/4 \pi)\bnabla \times \mathbf{B}$ the current density; $\gamma$ is a ratio of specific heats determined by an adiabatic equation of state. The magnetic field, $\mathbf{B}$, is divergence free: $\bnabla \cdot \mathbf{B} = 0$.
The magnetic diffusivity term in the induction equation is dropped, assuming a large magnetic Reynolds number in the solar interior, of order $R_m\sim 10^6$ \citep{Hood_2011}.
Hereafter, we resort to a background model which assumes the Sun to be spherically symmetric, non-rotating, non-magnetic, temporally stationary, capable of only sustaining adiabatic acoustic oscillations \citep[model S in][]{jcd}. We do not enlist the Poisson equation as we neglect perturbations to the gravitational potential $\phi$ --- that is, we use the Cowling approximation.

In steady state with no background flow ($\mathbf{v}_0 = \mathbf{0}$), the equilibrium force balance condition in model S is:
\begin{equation}
    \bnabla p_0 + \rho_0 \bnabla \phi = \mathbf{0}.
\end{equation}
A subscript `0' stands for zeroth order, static unperturbed fields.
We do not subscript $\phi$ (or the gravitational acceleration~$\mathbf{g}=-\bnabla\phi$) as the zeroth order gravitational potential (or field) is implied in the Cowling approximation.
When we perturb model S by introducing magnetic field $\Bv$, the steady-state equilibrium force-balance also includes the Lorentz-force due to the zeroth order magnetic field:
\begin{equation}
    \bnabla p_0 + \rho_0 \bnabla \phi = c^{-1}\,\mathbf{j}_0 \times \mathbf{B}_0.
\end{equation}

A departure from this steady state equilibrium, via the introduction of perturbative forces, causes the system to respond by exhibiting free oscillations $\xiv(\boldsymbol{r},t)$. In Fourier space, these free oscillations can be reconstructed from a superposition of ``normal-modes", labelled by an index~$k$, each of which have a characteristic frequency~$\omega_k$ and spatial pattern $\xiv_k(\boldsymbol{r})$ with excitation amplitudes~$a_k$:
\begin{equation}
    \xiv (\boldsymbol{r},t) = \sum_k a_k\,\xiv_k(\boldsymbol{r}) \exp(i \omega_k t).
\end{equation}

Introducing first-order, time-dependent perturbations as outlined in Appendix~\ref{sec: MHD_linearization}, we arrive at the linearized equation of motion for the above magnetohydrodynamic system:
\begin{equation} \label{eqn:total_wave_op}
    \rho_0 \,\omega_k^2\, \xiv_k = \cL_0 \xiv_k + \delta \cL \xiv_k,
\end{equation}
where $\cL_0$ is the linear acoustic wave operator for model-S while the operator $\delta \cL$ captures the forcing due to Lorentz-stresses. We solve Eqn.~(\ref{eqn:total_wave_op}) based on a perturbative analysis that involves the following two steps.
\begin{enumerate}
    \item We start with the unmagnetized case by setting $\mathbf{B}_0 = \mathbf{0}$ and solve the eigenvalue problem $\rho_0 \,\omega_{k,0}^2\, \xiv_{k,0} = \cL_0 \xiv_{k,0}$. This gives us the eigenfrequencies $\omega_{k,0}$ and eigenmodes $\xiv_{k,0}$ associated with the unperturbed model-S.
    \item We introduce a non-zero background magnetic field $\mathbf{B}_0$ and therefore ``switch-on" the $\delta \cL$ part of the wave operator in Eqn.~(\ref{eqn:total_wave_op}). Consequently, to accommodate for the changes due to $\delta \cL$, we introduce perturbed eigenfrequencies $\omega_k = \omega_{k,0} + \delta \omega_k$ and eigenfunctions $\xiv_k = \sum_{k'} c_{k'}\, \xiv_{k',0}$. Note that we neglect the distortions in the solar structure induced due to the presence of magnetic fields. Accounting for this would involve mapping each point $\mathbf{r}$ on the model S spherical Sun onto a point $\mathbf{x}$ on the distorted Sun \citep{goughmag}.
\end{enumerate}

The linearized equation of motion (see derivation of Eqn.~\ref{eqn: linearized_MHD}) for the unmagnetized case is given by \citep[also refer to][]{jcd_notes}: 
\begin{equation} \label{eqn:sol_wave_eqn}
    \mathcal{L}_0 \xiv_{k,0}  = - \grad (\rho_0 c_s^2\, \bnabla \cdot \boldsymbol{\xi}_{k,0} - \rho_0 g\, \boldsymbol{\xi}_{k,0} \cdot \ev{r}) - g \,\ev{r} \bnabla \cdot (\rho_0\, \boldsymbol{\xi}_{k,0}) = \rho_0 \,\omega_{k,0}^2\, \xiv_{k,0},
\end{equation}
where $c_s(r), \rho_0(r)$, and $g(r)$ are the sound speed, density, and gravity (directed radially inward) respectively, and $\bnabla$ denotes the covariant spatial derivative operator. For all ensuing calculations and derivations, we write Eqn.~(\ref{eqn:sol_wave_eqn}) in the form $\mathcal{L}_0 \boldsymbol{\xi}_k = \rho_0 \omega_k^2 \boldsymbol{\xi}_k$, where the magnetically unperturbed wave operator $\mathcal{L}_0$ is self-adjoint \citep{goedbloed2004}. To solve for the eigenmodes of the unperturbed model S, the boundary conditions employed are: (a) $\xiv$ and the Eulerian pressure perturbation at $r=0$ is finite, and (b) the Lagrangian pressure perturbation at $r=R_{\odot}$ vanishes \citep[see Section 17.6 in][]{Cox1980_Book}. As already mentioned earlier, the Cowling approximation is used and hence the gravitational Poisson equation is not needed while finding the eigenmodes. We suppress the subscript `0' in the unperturbed eigenfunctions $\xiv_{k,0}$ and eigenfrequencies $\omega_{k,0}$ for the rest of this paper. Unless specified otherwise, any instance of $\omega_k$ or $\xiv_k$ should be assumed to imply eigenfrequencies and eigenfunctions of Eqn.~(\ref{eqn:sol_wave_eqn}), respectively.
The Sun is treated as a fluid body with vanishing shear modulus and hence is unable to sustain shear waves (although the presence of magnetic fields complicates this assumption). Thus, the eigenfunctions of the background model contain no toroidal components \citep[see Chapter 8 of][]{DT98}, rendering them purely spheroidal. We write the displacement field $\boldsymbol{\xi}(\boldsymbol{r})$ in the basis of vector spherical harmonics (and thereafter GSH) as follows
\begin{eqnarray} 
    \boldsymbol{\xi}(r,\theta,\phi) &=& \sum_{n,\ell,m} {}_nU{}_{\ell}(r) \,Y_{\ell m}(\theta,\phi) \,\ev{r}
 + {}_nV{}_{\ell}(r) \, \grad_1 Y_{\ell m}(\theta,\phi) \label{eqn: xi_exp} \\
     &=& \sum_{n,\ell,m}  {}_n\xi{}_{\ell}^-(r) \,Y_{\ell m}^-(\theta,\phi) \,\ev{-} + {}_n\xi{}_{\ell}^0(r) \,Y_{\ell m}^0(\theta,\phi) \,\ev{0} + {}_n\xi{}_{\ell}^+(r) \,Y_{\ell m}^+(\theta,\phi) \,\ev{+}. \label{eqn: xi_exp_GSH}
 \end{eqnarray}
 Here, $\boldsymbol{r} = (r,\theta,\phi)$ denote spherical polar coordinates, with basis vectors $(\hat{e}_r,\hat{e}_{\theta},\hat{e}_{\phi})$ and $k = (n,\ell,m)$ where $n$ is the radial order, $\ell$ the angular degree, and $m$ the azimuthal order. The dimensionless lateral covariant derivative operator is denoted by $\boldsymbol{\grad}_1 = \ev{\theta}\,\partial_{\theta} + \ev{\phi}\,(\sin\theta)^{-1}\partial_{\phi}$. The basis vectors in spherical polar coordinates are related to those in the GSH basis via
 \begin{equation}
     \ev{-} = \frac{1}{\sqrt{2}}(\ev{\theta} - i \ev{\phi}), \qquad \ev{0} = \ev{r}, \qquad \ev{+} = -\frac{1}{\sqrt{2}}(\ev{\theta} + i \ev{\phi}).
 \end{equation}
 
 The vanishing toroidal component in~(\ref{eqn: xi_exp}) imposes the constraint that $\xi_{k}^{-} = \xi_{k}^{+}$ \citep[Appendix C of][]{DT98}. Owing to $\cL_0$ being self-adjoint, the eigenvalues $\omega_k$ of the unperturbed state are real and the eigenfunctions $\boldsymbol{\xi}_k$ corresponding to distinct eigenvalues are orthogonal. For convenience, $\boldsymbol{\xi}_k$ is normalized to satisfy the orthonormality condition
\begin{equation} \label{eqn: orthonormality}
  \solintv \rho\,\boldsymbol{\xi}_{k'}^* \cdot \boldsymbol{\xi}_{k} = \delta_{n'n}\, \delta_{\ell' \ell}\, \delta_{m' m}.
\end{equation}


Next, as derived in Appendix~\ref{sec: MHD_linearization} (see Eqn~[\ref{eqn: pert_linearized_MHD}]), we introduce a magnetically induced perturbation through the operator $\delta \cL$ where,
\begin{eqnarray}
    4\pi\,\delta \cL \xiv =  \mathbf{B}_0 \times (\bnabla \times \mathbf{B}_1) - (\bnabla \times \mathbf{B}_0) \times \mathbf{B}_1 - \bnabla [\xiv \cdot (\mathbf{j}_0 \times \mathbf{B}_0)].
\end{eqnarray}
$\mathbf{B}_0$ is the zeroth-order background magnetic field and $\mathbf{B}_1 = \bnabla \times (\xiv \times \mathbf{B}_0)$ is the first order perturbation. Henceforth, we shall drop the subscript `0' for the zeroth-order magnetic field. Therefore, any instance of $\Bv$ hereafter shall denote the zeroth-order magnetic field, unless specified otherwise. Like $\xiv$ (in Eqn~[\ref{eqn: xi_exp_GSH}]), we also expand the magnetic field and the corresponding Lorentz stress using the basis of generalized spherical harmonics~$\boldsymbol{Y}_{st}^{N}$
 \begin{eqnarray}
     \mathbf{B}(r,\theta,\phi) &=& \sum_{s=0}^{\infty} \sum_{t=-s}^{s}\sum_{\mu} B_{st}^{\mu}(r)\, Y_{st}^{\mu}(\theta,\phi) \,\ev{\mu} \label{eqn: B_exp_GSH},\\
     \boldsymbol{\mathcal{H}}(r,\theta,\phi) &=& \sum_{s=0}^{\infty} \sum_{t=-s}^{s} \sum_{\mu\nu} h_{st}^{\mu \nu}(r)\, Y_{st}^{\mu + \nu}(\theta,\phi)\, \ev{\mu}\, \ev{\nu}, \label{eqn: H_exp}
 \end{eqnarray}
where in Eqns.~(\ref{eqn: B_exp_GSH}) and~(\ref{eqn: H_exp}) $\mu$ and $\nu$ take values $-1$, 0, $+1$. In practice, the angular degree $s$ is usually truncated at some desired upper limit $s_{\rm{max}}$. Invoking the divergence-free condition on the magnetic field, $\bnabla \cdot \Bv = 0$, we have $B_{st}^{+} + B_{st}^{-} = \partial_r (r^2 B^{0}_{st})/(r \Omega_s^0)$, where $\Omega_{\ell}^{\pm N} = \sqrt{\frac{1}{2}(\ell \pm N)(\ell \mp N + 1)}$. The realness of $\mathbf{B}$ further imposes the constraint $B_{st}^{\mu\,*} = (-1)^t B_{s\bar{t}}^{\bar{\mu}}$, where overbars represent negatives (i.e., ${\bar t} = -t$), a property that follows from its realness condition. An equivalent constraint on the Lorentz stress $\boldsymbol{\mathcal{H}}$, which accounts for a solenoidal $\Bv$ field, seems unlikely. Therefore, as elucidated in Appendix~\ref{sec: H_expr}, we begin constructing our $\boldsymbol{\mathcal{H}}$ by choosing a divergence-free $\Bv$ field. Eqn.~(\ref{eqn:h_B_relation}) relates components of $\cH$ to $\Bv$. $\cH$, by construction, satisfies the symmetry property $h^{\mu\nu}_{st} = h^{\nu\mu}_{st}$ ($\because \cH = \Bv\Bv$), and $h^{\mu\nu\,*}_{st} = (-1)^t h_{s\bar{t}}^{\bar{\mu}\bar{\nu}}$. The introduction of magnetic perturbations introduces a set of surface terms which can be found in Eqn.~(\ref{eq: mag_surface_terms}). Further details about the importance of these terms and why their contribution is negligible for the Model S \citep{jcd} can be found in Appendix \ref{sec: deriv_cpmat}.

\subsection{Forward problem} \label{sec: Theory2}
The eigenfrequencies of the unmagnetized state (eqn~[\ref{eqn:sol_wave_eqn}]) are denoted by ${}_n\omega{}_{\ell}$. They are independent of the azimuthal order~$m$ because of the spherical symmetry of the background model \citep[we use model S of][]{jcd_evol}. Therefore, each multiplet ${}_n\mathrm{S}_\ell$ is $2\ell +1$ degenerate.
The introduction of (non-spherically symmetric) perturbations breaks this degeneracy and gives rise to frequency splitting. The aim of our forward problem is to calculate the perturbed eigenfrequencies under the action of Lorentz stresses. In calculating frequency shifts of an erstwhile degenerate multiplet ${}_{n_0}\mathrm{S}_{\ell_0}$, we apply the formalism of quasi-degenerate perturbation theory.
Therefore, perturbations in the eigenfunctions and eigenfrequencies of a target multiplet ${}_{n_0}\mathrm{S}_{\ell_0}$ are assumed to arise from eigenmodes whose unperturbed eigenfrequencies $\omega_{k}$ lie within the range $[\omega_{\rm{ref}} - \Delta \omega, \omega_{\rm{ref}} + \Delta \omega]$, where the reference angular frequency~$\omega_{\rm{ref}}$ is chosen to be ${}_{n_0}\omega_{\ell_0}$, and $\Delta \omega$ defines a window in frequency around $\omega_{\rm{ref}}$.
Any mode with frequency $\omega_{k}$ outside the range~$[\omega_{\rm{ref}} - \Delta \omega, \omega_{\rm{ref}} + \Delta \omega]$ is assumed to not ``talk" to multiplet~${}_{n_0}\mathrm{S}_{\ell_0}$. Carrying out the standard quasi-degenerate perturbation analysis discussed in Appendix~\ref{sec: Pert_deriv}, frequency splitting within each multiplet is given by the following expression:
\begin{equation} \label{eqn:evalue_problem_QDPT}
    \sum_k [ \Lambda_{k'k} - (\omega_{\rm{ref}}^2 - \omega_k^2) \,\delta_{k'k} ] c_k =  \delta\omega^2 \, c_{k'},
\end{equation}
where the matrix $\Lambda_{k'k} = \solintv\, \boldsymbol{\xi}_{k'}^*\cdot \delta \mathcal{L} \boldsymbol{\xi}_{k} = \braket{{\boldsymbol{\xi}_{k'}} | \delta \mathcal{L} \boldsymbol{\xi}_{k}}$ is called the coupling matrix and $\delta\omega^2 = 2\, \omega{}_{\mathrm{ref}}\,\delta \omega$. Here, $\delta \omega$ represents the shift in eigenfrequencies about the reference frequency $\omega_{\mathrm{ref}}$. The shifted eigenfrequencies, in general, do not carry an unperturbed mode index $k'$ because $n',\ell',m'$ cease to behave as good quantum numbers. For conciseness, it is common to define $\mathcal{Z}_{k'k} = \Lambda_{k'k} - (\omega_{\rm{ref}}^2 - \omega_k^2)\, \delta_{k'k}$ as the supermatrix. The degenerate eigenmode with unperturbed frequency ${}_n\omega{}_{\ell}$ is denoted by the triplet~$k=(n,\ell,m)$. The derivation of these matrix elements is outlined in Appendix~\ref{sec: deriv_cpmat}. The simplified form of the coupling matrix is written as a radial integral of $h_{st}^{\mu\nu}(r)$ components weighted by sensitivity kernels ${}_{k'k}\cB_{st}^{\mu\nu}(r)$ according to
\begin{align}
    \Lambda_{k'k} &= \sum_{st} \sum_{\mu\nu} \solint r^2{}_{k'k}\cB_{st}^{\mu\nu} (r) \,h_{st}^{\mu\nu} (r) \label{eqn:lamda_decomp}\\
    &= \sum_{s,t} \solint r^2 \,\{{}_{k'k}\cB_{st}^{00} \,h_{st}^{00} + {}_{k'k}\cB_{st}^{++}\,[h_{st}^{--} (-1)^{\ell'+\ell+s} + h_{st}^{++}] + 2\, {}_{k'k}\cB_{st}^{0+}\,[h_{st}^{0-}(-1)^{\ell'+\ell+s} + h_{st}^{0+}] + 2\,{}_{k'k}\cB_{st}^{+-}\,h_{st}^{+-}\} \label{eqn: lambda_m}.
\end{align}
Explicit expressions and symmetry relations for corresponding sensitivity kernels ${}_{k'k}\cB^{\mu \nu}_{st}$ are stated in Appendix~\ref{sec:mag_kern}. Hereafter, we shall drop the mode subscripts $k'k$ in ${}_{k'k}\cB^{\mu \nu}_{st}$ for notational simplicity. Expressions for Lorentz-stress components $h^{\mu \nu}_{st}$ may be found in Appendix~\ref{sec: H_expr}. In writing Eqn.~(\ref{eqn: lambda_m}) we use the symmetry in the Lorentz-stress tensor-expansion components, $h^{\mu \nu}_{st} = h^{\nu \mu}_{st}$, and the fact that there are only four independent components of $\cB_{st}^{\mu\nu}$. The eigenvalues $\delta \omega^2$ of the supermatrix $\mathcal{Z}_{k'k}$ are given by Eqn.~(\ref{eqn:evalue_problem_QDPT}). These relate to $\Omega_p$, the perturbed frequencies, via $\Omega_p = (\omega_{\rm{ref}}^2 + \delta\omega^2_p)^{1/2}$ where $p$ denotes a new set of labels different from the unperturbed label $k$. In the absence of spherical or azimuthal symmetry and cross-mode coupling, $n,\ell,m$ are no longer `good' quantum numbers and therefore inhibit the direct mapping of the perturbed modes with the erstwhile unperturbed modes. Note that for the purpose of inverse problems, the proper method of parameterizing the inversion is to use $\cH/\rho$ and corresponding effective kernels $\rho \cB$. This is discussed in further detail in Appendix \ref{sec: deriv_cpmat}. The forward problem of Eqn.~(\ref{eqn: lambda_m}) remains unchanged under this consideration. 

\subsection{Coupled multiplets}\label{sec: coupled_multiplets}

A general spheroidal p-mode wavefield in the spectral domain can be expanded in the basis of the unperturbed spheroidal p-modes as $\xi(\boldsymbol{r,\omega}) = \sum_{k}\varphi_k(\omega)\xiv_k(\boldsymbol{r})$. Following the concise notation used in \cite{hanasoge17_etal}, we henceforth write $\varphi_k(\omega)$ as $\varphi_k^{\omega}$,
which contains the phase and amplitude of mode $k$. The time series of the spherical harmonic oscillations at the solar surface (and hence $\varphi_k^{\omega}$) are supplied by the HMI and MDI missions. The change in this wavefield $\xiv$ in the presence of a perturbation can be expressed as
\begin{equation}
    \delta \xiv = \sum_k \delta \varphi_k^{\omega} \xiv_k.
\end{equation}

As shown in \cite{woodard16} and \cite{hanasoge17_etal}, the expectation value of the cross-spectral correlation signal can be modelled as:
\begin{equation} \label{eqn: mode_coupling}
    \langle \varphi_{k'}^{\omega'} \delta\varphi_k^{\omega *} + \delta \varphi_{k'}^{\omega'} \varphi_k^{\omega *} \rangle = (N_{k'} R_k^{\omega *} |R_{k'}^{\omega'}|^2 + N_{k} R_{k'}^{\omega'} |R_{k}^{\omega}|^2) \Lambda_{k'k} 
\end{equation}
where $\langle A \rangle$ represents the statistical expectation value of parameter $A$, $N_k$ is the mode-amplitude normalization and $R_k^{\omega} = (\omega^2 - \omega_k^2)^{-1}$.
The singularity at $\omega = \omega_k$ is avoided by introducing a tiny imaginary component to $\omega_k$ (which is representative of negligible attenuation). This results in damping of the mode. The damping rate $\gamma_k \sim 1/\tau$ where $\tau$ is the e-folding lifetime of the mode. The lifetime or damping rate can be accurately computed by modeling the power spectrum of modes in frequency and fitting a Lorenzian profile \citep{schou94}. If $\Delta$ is the FWHM then $\tau = (\pi \Delta)^{-1}$. The observed lifetime of modes in the Sun vary from days to a few months \citep{Korrenzik13_JOP}. Therefore, an estimation of $\langle \varphi_{k'}^{\omega'} \delta\varphi_k^{\omega *} + \delta \varphi_{k'}^{\omega'} \varphi_k^{\omega *} \rangle$ from observations allow us to recover elements in $\Lambda_{k'k}$.

The elements~$\Lambda_{k'k}$ thus obtained define a non-sparse matrix for a general non-axisymmetric field. This is unlike the $a$-coefficient formalism, which is restricted to an inverse problem for axisymmetric fields and therefore has a diagonal coupling matrix. 
Because $\Lambda_{k'k}$ has eigenvalues $\delta \omega^2 = 2 \omega_{\rm{ref}}\,\delta \omega$, where $\omega_{\rm{ref}}$ is a fiducial reference frequency (refer to Appendix~\ref{sec: Pert_deriv}). One can then choose to frame the following statement for an inverse problem \citep[Section 14.2.9 and 14.3.5]{DT98}:
\begin{equation}\label{eqn: cst}
    \Lambda_{k'k}/(2\omega_{\rm{ref}}) = \omega_{\rm{ref}} \sum_{s=0}^{\infty} \sum_{t = -s}^{s} c_{st}^{n'\ell'n\ell}\int_{\Omega} Y_{\ell'm'}^{0*} Y_{st}^0 Y_{\ell m}^0 \,\mathrm{d}\Omega ,
\end{equation}
where the coefficients~$c_{st}^{n'\ell'n\ell}$ remain to be determined. The solid angle integration is carried out over the complete surface of a sphere. The integral involving three spherical harmonics can be conveniently expressed in terms of Wigner 3-$j$ symbols as follows ~\citep[][C.225]{DT98}:
\begin{equation}
    \int_{\Omega} Y_{\ell'm'}^{0*} Y_{st}^0 Y_{\ell m}^0 \,\mathrm{d}\Omega = 4\pi (-1)^{m'} \gamma_{\ell'} \gamma_{s} \gamma_{\ell} \wigred{0}{0}{0} \wigred{-m'}{t}{m},
\end{equation}
where~$\gamma_\ell=\sqrt{(2\ell+1)/(4\pi)}$.
Upon comparing Eqn.~(\ref{eqn: cst}) with Eqn.~(\ref{eqn:lamda_decomp}), we find that
\begin{eqnarray}
    c_{st}^{n'\ell'n\ell} &=& \frac{1}{2 \omega_{\rm{ref}}^2} \sum_{\mu,\nu} \int \mathrm{d}r\,r^2 \cB_{st}^{\mu\nu} h_{st}^{\mu\nu} \Big/ \int_{\Omega} Y_{\ell'm'}^{0*} Y_{st}^0 Y_{\ell m}^0 \,\mathrm{d}\Omega \\
    &=& \frac{1}{2 \omega_{\rm{ref}}^2} \sum_{\mu,\nu} \int \mathrm{d}r\,r^2 \cG_{s}^{\mu\nu} h_{st}^{\mu\nu} \Big/ \wigred{0}{0}{0}, \label{eqn: inv_prob_cst}
\end{eqnarray}
where $\cB_{st}^{\mu\nu} = 4\pi (-1)^{m'}\gamma_{\ell'} \gamma_{s} \gamma_{\ell}\wigred{-m'}{t}{m}\cG_{s}^{\mu\nu}$ (Eqn.~[\ref{eqn: m_ind_kern}]).
As $\cG_{s}^{\mu\nu}$ does not have an $m$ or $m'$ dependence,
the coefficient $c_{st}^{n'\ell'n\ell}$ only depends on the multiplet labels --- $(n',\ell')$ and $(n,\ell)$.
Doing away with the $m$-dependence reduces the inverse problem in Eqn.~(\ref{eqn: cst}) for each submatrix of dimension $(2\ell'+1) \times (2\ell + 1)$ to that for a single element --- $c_{st}^{n'\ell'n\ell}$ --- as shown in Eqn.~(\ref{eqn: inv_prob_cst}). Again, while inverting for the components of the Lorentz stress tensor from $c_{st}^{n'\ell'n\ell}$, the effective kernels should be $\rho \cG_{s}^{\mu\nu}$ for obtaining $h_{st}^{\mu\nu}/\rho$. The $c_{st}^{n'\ell'n\ell}$ are the so-called structure coefficients. It is commonplace in terrestrial seismology to use the structure coefficients to construct the `generalized splitting functions'
\begin{equation}
    \eta^{n'\ell'n\ell}(\theta,\phi) = \sum_{s=|\ell'-\ell|}^{\ell'+\ell}\sum_{t=-s}^{s} c_{st}^{n'\ell'n\ell} \,Y_{st}(\theta,\phi)
\end{equation}
These splitting functions allow for a convenient form of visualization of the splitting due to self-coupling or cross-coupling of multiplets under a generic non-axisymmetric structure perturbation (the Lorentz-stress field in our case). As mentioned in \cite{DT98}, the function $\eta(\theta,\phi)$ at any location $(\theta,\phi)$ on the surface illustrates the local radially averaged internal 3D structure, as ``seen" through the kernels. A demonstrative plot of splitting functions due to the self-coupling of mode $\mode{2}{8}$ and its cross-coupling with mode $\mode{3}{7}$ is shown in Figure~\ref{fig:splitting_functions}.

It may be noted that $\eta^{n'\ell'n\ell}(\theta,\phi)$ is purely real. This can be verified by using (a) the kernel property $\cG_{s}^{\mu\nu} = (-1)^{\ell'+\ell+s} \cG_s^{\bar{\mu}\bar{\nu}}$ with $\ell'+\ell+s = $~even because of the selection rule in Eqn.~(\ref{eqn: inv_prob_cst}), (b) realness of $\cH$, i.e., $h_{st}^{\mu\nu} = (-1)^t h_{s\bar{t}}^{\bar{\mu}\bar{\nu}}$, and (c) the GSH property $Y_{st}^{N*} = (-1)^{N+t} Y_{s\bar{t}}^{\bar{N}}$.

\subsection{Isolated-multiplet approximation} \label{sec: isolated_multiplet}

The extent of coupling between two multiplets depends primarily on the proximity of their unperturbed frequencies, i.e., $|{}_{n_1}\omega_{\ell_1} - {}_{n_2}\omega_{\ell_2}|$. The more separated they are in frequency, the weaker is the coupling. As shown in equation (143) of \cite{lavely92} this separation and hence the importance of considering cross-coupling can be quantified through $\nu_{\mathrm{CC}}$ the ``coupling strength coefficient". When a multiplet ${}_{n}\mathrm{S}_{\ell}$ is well separated ($\nu_{CC} < 1nHz$) from any other multiplet in the frequency domain, contribution due to modes in other multiplets become significantly weak and therefore ${}_{n}\mathrm{S}_{\ell}$ can potentially be treated as isolated in frequency. Under such circumstances one can use what is known as the ``isolated-multiplet" approximation where only the modes with different azimuthal orders $m$ belonging to the same multiplet ${}_{n}\mathrm{S}_{\ell}$ are considered for perturbation analysis. In such cases, $\omega_{\rm{ref}} = {}_n\omega{}_{\ell}$ and Eqn.~(\ref{eqn:evalue_problem_QDPT}) reduces to 
\begin{equation} \label{eqn:evalue_problem_DPT}
    \sum_{m'} \Lambda_{mm'}\, c_{m'} = \delta\omega^2 \,c_m.
\end{equation}
Note the change in the index from $k=(n,\ell,m)$ to $m$. This is because we restrict the coupling to a specific multiplet and therefore $n$ and $\ell$ remain good quantum numbers whereas $m$ does not.

\subsubsection{$a$-Coefficients} \label{sec: a-coefficients}
If we further assume the perturbation to be axisymmetric, even $m$ remains a good quantum number. This is because the coupling matrix $\Lambda_{mm'} \sim \delta_{mm'}$ owing to the selection rule imposed by the Wigner-$3j$ symbols.
In that case,
Eqn.~(\ref{eqn:evalue_problem_DPT}) gives us the frequency shifts corresponding to each mode $(n,\ell,m)$. We divide this by twice the degenerate frequency ${}_n\omega{}_{\ell}$ corresponding to that multiplet to determine the $2\ell+1$ split frequencies $\delta {}_n\omega{}_{\ell m}$.
Finally, we obtain perturbed frequencies ${}_n\omega{}_{\ell m} = {}_n\omega{}_{\ell} + \delta {}_n\omega{}_{\ell m}$. It is customary to represent such frequency splittings (or equivalently frequency shifts) by so-called `$a$-coefficients',
\begin{equation} \label{eqn: a-coeffs}
\delta {}_n\omega{}_{\ell m} = \sum_{j=0}^{j_\text{max}} a^{n\ell}_j\, \cP_j(m),
\end{equation}
where $a_{j}^{n\ell}$ are the $a$-coefficients and $\cP_j(m)$ are a set of orthogonal basis polynomials in $m$ of degree $j$. Eqn.~(\ref{eqn: a-coeffs}) therefore represents a fitting of frequency splittings $\delta{}_n\omega{}_{\ell m}$ using appropriate weights $a_{j}^{n\ell}$ on an orthogonal basis \citep{schou_pol_94}.

Axisymmetry implies that the azimuthal order $t = 0$ in $h_{st}^{\mu\nu}$, and therefore all other components of the Lorentz-stress tensor except $h_{s0}^{\mu \nu}$ vanish. The selection rules imposed due to the Wigner-$3j$ symbols in the expressions for sensitivity kernels in Appendix~\ref{sec:mag_kern} in turn implies $m = m'$. This makes all off-diagonal elements vanish in the coupling matrix $\Lambda_{m'm}$ and, according to Eqn.~(\ref{eqn:evalue_problem_DPT}), its eigenvalues are given by its diagonal elements $\Lambda_{mm}$, i.e., $\delta{}_n\omega{}_{\ell m} \approx \Lambda_{mm}/\enc{2\,{}_n\omega{}_{\ell}}$. Using Eqn.~(\ref{eqn:lamda_decomp}) for $\Lambda_{mm}$, we have 
\begin{align}
\delta {}_n\omega{}_{\ell m} &= \frac{1}{2\,{}_n\omega{}_{\ell}}\sum_{s}\sum_{\mu\nu} \solint r^2\,\cB_{s0}^{\mu\nu} \,h_{s0}^{\mu\nu}\\
&=  \sum_{s} \left[\sum_{\mu\nu}\solint r^2 \mathcal{K}_{s}^{\mu\nu} h^{\mu\nu}_{s0} / \enc{2\,{}_n\omega{}_{\ell}}\right] \cP_{s}(m),
\end{align}
where $\cK^{\mu\nu}_s$ is the $m$-independent part of $\cB^{\mu\nu}_{s0}$ (see Appendix~\ref{sec:mag_kern}, Eqs.~[\ref{eqn:kern_m_dep}],~[\ref{eqn:wigner_P_ljm_relation}], and~[\ref{eqn:Kernel_m_separation}]). We identify the term in square brackets as being precisely the $a$-coefficients (see Eq.~[\ref{eqn: a-coeffs}]) of the axisymmetric splitting, and hence we assert
\begin{align}
    a_{s}^{n \ell} &= \solint r^2 \sum_{\mu,\nu} \mathcal{A}_{s}^{\mu\nu} \,h_{s0}^{\mu\nu} \label{eqn:a_coeffs_invprob1} \\
    &= \solint r^2 \{\cA_{s}^{00} \, h_{s0}^{00} + \cA_{s}^{++}\,[h_{s0}^{--} (-1)^{s} + h_{s0}^{++}] + 2\, \cA_{s}^{0+}\,[h_{s0}^{0-}(-1)^{s} + h_{s0}^{0+}] + 2\,\cA_{s}^{+-}\,h_{s0}^{+-}\}, \label{eqn:a_coeffs_invprob_expanded}\\
\end{align}
 where
\begin{equation}
    \cA_{s}^{\mu\nu} (r)= \cK^{\mu\nu}_{s}(r)/(2\,{}_n\omega{}_{\ell}).
    \label{eqn:a_coeff_kern}
\end{equation} 
The kernels $\cA^{\mu\nu}_s(r)$ represent $\cH$ sensitivities of $a$-coefficients. Eqn.~(\ref{eqn:a_coeffs_invprob1}) shows that each $a$-coefficient corresponding to a particular degree $s$ is proportionally dependent on the Lorentz-stress component of the same harmonic degree. As mentioned earlier as well as in Appendix \ref{sec: deriv_cpmat}, we emphasize that for an inverse problem, the parameter to invert for should be $h_{st}^{\mu\nu}/\rho$ and the effective kernels should be $\rho\cA_s^{\mu\nu}$. The formalism of $a$-coefficients is the same as that used in earlier similar studies \citep{Antia2000,baldner10} of estimating the effect of magnetic field on frequency splitting. \cite{Antia2000}, for example, sought to explain the statistically significant residual even $a$-coefficients after removing first and second order effects due to rotation by using either a 300kG field at the tachocline or using a 20kG field at $0.96R_{\odot}$. In this study, instead of trying to explain the observed $a$-coefficients with a choice of magnetic field, we suggest the inverse problem in Eqn.~(\ref{eqn:a_coeffs_invprob_expanded}).

However, we note  that the one-to-one linear relation between splitting coefficients and corresponding $\cH$ components arose as a consequence of the axisymmetry of the problem. A non-axisymmetric perturbative magnetic field does not accommodate a definition of the splitting-coefficient sensitivity kernels in as straightforward a manner as in Eqn.~(\ref{eqn:a_coeffs_invprob1}). This is because the frequency shifts are given by eigenvalues of the coupling matrix $\Lambda_{m'm}$, and eigenvalues of a matrix with nonzero off-diagonal terms are not, in general, linearly dependent on its matrix entries.

\subsubsection{Structure coefficients $c_{st}$} \label{sec: c_st_isolmultiplet}

Although the $a$-coefficient formalism does not accommodate the inversion of the components of a non-axisymmetric field from the splitting of an isolated multiplet, the formalism of structure coefficients outlined in Section~\ref{sec: coupled_multiplets} permits such an inversion. In this case, the coupling matrix is $(2\ell+1)\times(2\ell+1)$ and Eqn.~(\ref{eqn: cst}) can be written as
\begin{equation}
     \Lambda_{kk}/(2\omega_{k}) = \omega_{k} \sum_{s,t} c_{st}^{n\ell n\ell}\int_{\Omega} Y_{\ell m'}^{0*} Y_{st}^0 Y_{\ell m}^0 \,\mathrm{d}\Omega.
\end{equation}
Henceforth, for self-coupling we denote the structure coefficients as $c_{st}^{n\ell}$,
dropping superfluous labels in the superscript. The gaunt integral $\int_{\Omega} Y_{\ell m'}^{0*} Y_{st}^0 Y_{\ell m}^0 \,\mathrm{d}\Omega \propto \wigfull{\ell}{s}{\ell}{0}{0}{0}$. This induces the selection rules: (a) $s=$ even, and (b) $0 \leq s \leq 2 \ell$, thus allowing for the inversion of $c_{st}^{n\ell}$ for $s=0,2,4,6,...,2\ell$. As described before, the structure coefficients $c_{st}^{n\ell}$ can be used to compute the splitting function 
\begin{equation}
    \eta(\theta,\phi) = \sum_{s = 0}^{2\ell}\sum_{t=-s}^{s} c_{st}^{n\ell}\, Y_{st} (\theta,\phi)
\end{equation}
Here summation is over even $s$. A plot demonstrating splitting functions due to self-coupling of mode $\mode{2}{8}$ is shown in Figure~\ref{fig:splitting_functions}.





\section{Results} \label{Sec: Results}

We first analyse whether or not, in the presence of a general magnetic field, a multiplet ${}_{n}\mathrm{S}_{\ell}$ can be treated as isolated. 
To accomplish this, we choose a collection of multiplets with $0 \leq n \leq 20$ and $0 \leq \ell \leq 30$. Identifying `isolated' multiplets simplifies the problem immensely without having to resort to a computationally intensive cross-coupling calculation. This is because the forward problem involves solving an eigenvalue equation whose computational cost depends on the number of modes accounted for. In order to calculate the frequency shifts of a certain target multiplet ${}_n\rm{S}_{\ell}$, cross-coupling becomes important as the proximity in frequency to other multiplets ${}_{n'}\rm{S}_{\ell'}$ increases. It then becomes necessary to consider all the modes within a certain frequency window while performing a quasi-degenerate perturbation analysis. However, as the number of modes participating in the analysis increases, the coupling matrix $\Lambda_{k'k}$, and hence the supermatrix $\mathcal{Z}_{k'k}$, becomes increasingly large. This increases the computational burden, especially as one moves to larger values of $\ell$ (because each multiplet ${}_{n}\rm{S}_{\ell}$ contains $2 \ell +1$ singlets). Therefore, identifying where the isolated-multiplet assumption holds is a useful task, allowing one to carry out a degenerate perturbation analysis. 

On establishing the existence of such isolated multiplets, we analyze the kernels for $a$-coefficients (Eqn.~\ref{eqn:a_coeff_kern}) and highlight the various selection rules that emerge from the kernel expressions. Lastly, we discuss self-coupling under axisymmetric fields and carry-out the forward problem of finding the $a$-coefficients for an analytically constructed simple and yet realistic magnetic field.

\subsection{Mode classification} \label{sec: mode_labelling}

We use a $50 \,\mu$~Hz~frequency window for the quasi-degenerate perturbation analysis. It can be shown that differential rotation induces a coupling strength coefficient $\nu_{CC} \approx$ $1nHz$ between the interacting modes within a frequency window of $30 \,\mu$~Hz \citep{lavely92}. Differential rotation is the strongest perturbation over the SNRNMAIS Sun. Therefore, all multiplets capable of causing a $\nu_{CC} \geq 1nHz$  due to magnetic perturbations, lie within a window of $\Delta \omega = 50 \,\mu$~Hz. Therefore, any mode ${}_{n'}\rm{S}_{\ell'}$ that has an unperturbed frequency ${}_{n'}\omega{}_{\ell'}$ in the window ${}_n\omega{}_{\ell} - \Delta \omega \leq {}_{n'}\omega{}_{\ell'} \leq {}_n\omega{}_{\ell} + \Delta \omega$ is considered to couple non-trivially with ${}_{n}\rm{S}_{\ell}$ and hence is included in the eigenspace of modes considered for cross-coupling. We refer to this eigenspace as $\mathcal{K}$. We consider a Lorentz stress-field consisting of angular degrees $s=\{0,1,2,3,4,5,6\}$ containing all the azimuthal orders $t$ for a given $s$. Therefore, modes with $|\ell'-\ell| \leq 6$ are coupled. A uniform-strength Lorentz-stress field (varying as a function of $r$) is allocated to all the spatial wavenumber channels \begin{equation}\label{eqn: all_st_equal}
    h_{st}^{\mu\nu}(r) = h_0(r) \qquad \forall s,t .
\end{equation}
Here, $h_0(r) = b_0^2(r)$, with $b_0(r)$ being the profile of the `total' field shown in Figure~\ref{fig: field_strength_plot}. In addition, we impose the realness condition of $\cH$\,, $h^{\mu\nu\,*}_{st} = (-1)^t h_{s\bar{t}}^{\bar{\mu}\bar{\nu}}$\,, and the symmetry $h_{st}^{\mu\nu} = h_{st}^{\nu\mu}$. In Figure~\ref{fig: field_strength_plot}, the field strength $b_0(r)$ as a function of radius is inspired from the references cited in Section~\ref{sec: Theory3}. Allocating this to all the components $h_{st}^{\mu\nu}(r)$ certainly guarantees a general magnetic field with a total strength significantly exceeding current estimates of the solar internal field strength. Therefore, in the ensuing calculation, the extent of mode-coupling should be an over-estimation. Thus, modes labelled `coupled' could, in reality, be `isolated' but modes labelled `isolated' are expected to be good for carrying out degenerate perturbation analysis.

\begin{figure}[t] 
\centering
\includegraphics[width=1.03\linewidth]{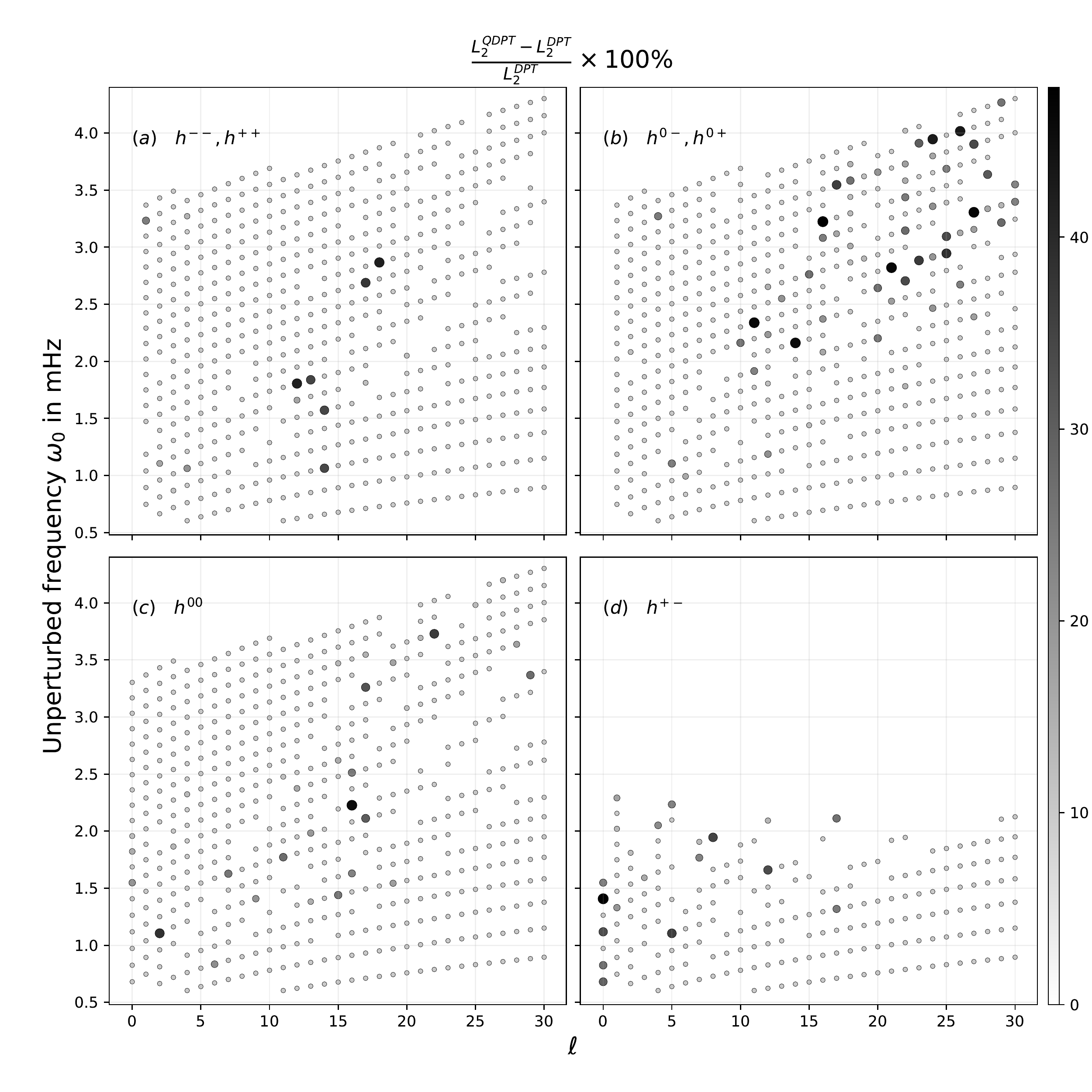} 
\caption{The relative offset of $L_2^\text{QDPT}$ as compared to that of $L_2^\text{DPT}$ (see Eqns.~\ref{eqn: l2_Q} and~\ref{eqn: l2_D}) due to a general Lorentz-stress field $\cH = \sum_{s,t} h_{st}^{\mu\nu}\,Y_{st}^{\mu+\nu}\, \ev{\mu}\, \ev{\nu}$ where $h_{st}^{\mu\nu} = h_0(r)$, where $h_0(r)$ is the `total' field strength shown in Figure~\ref{fig: field_strength_plot}. The effect arising from the presence of four independent components of $h_{st}^{\mu\nu}$ is plotted in the four panels. The gray-scale intensity and size of each `o' (representing a multiplet) increases with increasing departure of $\delta{}_{n}\omega{}_{\ell m}^\text{Q}$ from $\delta {}_{n}\omega{}_{\ell m}^\text{D}$. A larger and darker `o', implies stronger cross coupling for that multiplet.} 
\label{fig: mode_labelling}
\end{figure}

In Figure~\ref{fig: mode_labelling} we color-code multiplets to indicate the departure of frequency shifts obtained from quasi-degenerate perturbation $\delta{}_{n}\omega{}_{\ell m}^{Q}$ as compared to shifts obtained from a degenerate perturbation $\delta{}_{n}\omega{}_{\ell m}^{D}$. As the perturbations are not axisymmetric, the azimuthal order does not stay well defined after coupling and hence prevents a one-one mapping of unperturbed to perturbed modes. This hinders an explicit comparison of frequency shifts on a singlet-by-singlet basis. Hence, to quantify the departure of $\delta{}_{n}\omega{}_{\ell m}^{Q}$ from $\delta{}_{n}\omega{}_{\ell m}^{D}$ we calculate the Frobenius norm of these frequency shifts:
\begin{eqnarray}
    L_2^\text{QDPT} &=& \sqrt{\sum_m (\delta {}_{n}\omega{}_{\ell m}^\text{Q})^2} \qquad \text{for cross-coupling,} \label{eqn: l2_Q}\\
    L_2^\text{DPT} &=&  \sqrt{\sum_m (\delta {}_{n}\omega{}_{\ell m}^\text{D})^2} \qquad \text{for self-coupling.} \label{eqn: l2_D}
\end{eqnarray}
The gray-scale intensity in Figure~\ref{fig: mode_labelling} indicates the relative offset of $L_2^\text{QDPT}$ as compared to $L_2^\text{DPT}$ for a mode ${}_n\rm{S}_{\ell}$ marked as an `o'. Larger offset indicates stronger relative offset and are marked as a bigger and darker `o'. For visual clarity, we have set a lower threshold on the size of the `o' --- any mode with less than $5\%$ offset has the same size. Certain ${}_n\rm{S}_{\ell}$ have unperturbed frequencies proximal to adjacent multiplets ${}_{n'}\rm{S}_{\ell'}$ prior to perturbation. Depending on the magnitude of perturbation and its sensitivity, modes of such ${}_n\rm{S}_{\ell}$ become strongly entangled with adjacent modes. Because the labels $n,\ell,m$ lose meaning after perturbation, it is impossible to disentangle modes belonging to our target multiplet from those belonging to its adjacent multiplets. For such multiplets we cannot obtain $\delta {}_{n}\omega{}_{\ell m}^\text{Q}$ and therefore avoid plotting them in Figure~\ref{fig: mode_labelling}. We refer to these as `invisible modes' in the context of this figure. 

We plot four different cases, considering the effect of independent components of $\cH$ in each. This helps in explicitly demonstrating which multiplets have a non-trivial difference between $\delta {}_n\omega{}_{\ell m}^\text{Q}$ and $\delta {}_n\omega{}_{\ell m}^\text{D}$ under the influence of a particular Lorentz-stress component. For example, we plot the effect due to the presence of only $h_{st}^{00}$ or only $h_{st}^{+-}$ in Figure~\ref{fig: mode_labelling}(c) and Figure~\ref{fig: mode_labelling}(d) respectively. In Figure~\ref{fig: mode_labelling}(a), we plot the effect of the presence of $h_{st}^{--}$ and $h_{st}^{++}$ as these are coupled through the realness of $\cH$. Similarly for Figure~\ref{fig: mode_labelling}(b) we consider the presence of $h_{st}^{0-}$ and $h_{st}^{0+}$. Because we choose the same radial profile for all $h_{st}^{\mu\nu}$ (refer Eqn~[\ref{eqn: all_st_equal}]), differences in the four panels primarily reflect the effect of sensitivity kernels $\mathcal{B}_{st}^{\mu\nu}$ for changing $\mu,\nu$. We find the maximum number of invisible modes in Figure~\ref{fig: mode_labelling}(d) corresponding to $h_{st}^{+-}$.
This is because $\mathcal{B}_{st}^{+-}$ is at least an order of magnitude larger than other components (Figures~\ref{fig: kernels_1D} and \ref{fig: a2_kern_allcompare}), thereby causing multiplets to overlap in frequency and hindering post perturbation mode identification. $\mathcal{B}_{st}^{00}$ and $\mathcal{B}_{st}^{--}$ have comparable magnitudes, are smaller than $\mathcal{B}_{st}^{+-}$ but larger than $\mathcal{B}_{st}^{0-}$.
Therefore, the presence of $h_{st}^{00}$ (Figure~\ref{fig: mode_labelling}(c)) and $h_{st}^{--}, h_{st}^{++}$ (Figure~\ref{fig: mode_labelling}(a)) induce an intermediate number of invisible modes and $h_{st}^{0-},h_{st}^{0+}$ (Figure~\ref{fig: mode_labelling}(b)) induce the least number of invisible modes. It should be noted that $\mathcal{B}_{st}^{0-}$ is the weakest component of the sensitivity kernel and therefore we expect it to couple modes the least. However, the number of darker `o's due to the presence of $h_{st}^{0-},h_{st}^{0+}$ is more than the other cases because: (i) it contains the least number of invisible modes as many of its strongly coupled multiplets are invisible in the other components. Hence, many of the dark `o's in Figure~\ref{fig: mode_labelling}(b) are actually invisible in Figure~\ref{fig: mode_labelling}(a), (b), and (d), and (ii) the gray-scale intensity depends on the relative offset $(L_2^\text{QDPT}-L_2^\text{DPT})/L_2^\text{DPT}$. The lowest sensitivity of $h_{st}^{0-},h_{st}^{0+}$ causes the least frequency shifts and therefore $L_2^\text{DPT}$ is much smaller in magnitude, resulting in seemingly larger relative offsets.

Returning to the question we had sought to address earlier, because of the large number of small, white 'o' symbols across all the components, we can assert that there are multiplets which can be treated as isolated. However, because Lorentz-stress in the Sun comes as a combination of all the four cases, a multiplet can be regarded as free of cross-coupling only when it is isolated in all the four cases. Thus, although there are many isolated-multiplets in Figures~\ref{fig: mode_labelling}(a), (b), and (c) the number is restricted to those which are deemed isolated in (d). So, in the patch of $n,\ell$ we have chosen, we do find modes that can be treated as isolated. Therefore, in the following sections we explore the simplifications to the forward and inverse problem while restricting our analysis to these isolated-multiplets. However, as mentioned earlier, it should be noted that our choice of assigning the same strength to all $h_{st}^{\mu\nu}$ is supposed to cause an over-estimation of the coupling and therefore an under-estimation of the number of isolated-multiplets. A similar plot showing the validity of isolated-multiplet assumption when considering differential rotation as the sole perturbation can be found in Figure~\ref{fig:DR_coupling} under Appendix~\ref{sec: DR_mode_labelling}.

\subsection{Self-coupling due to an axisymmetric magnetic field.} \label{sec: self-coupling-axisymmetric}

\begin{figure}[t] 
\centering
\includegraphics[width=1.03\linewidth]{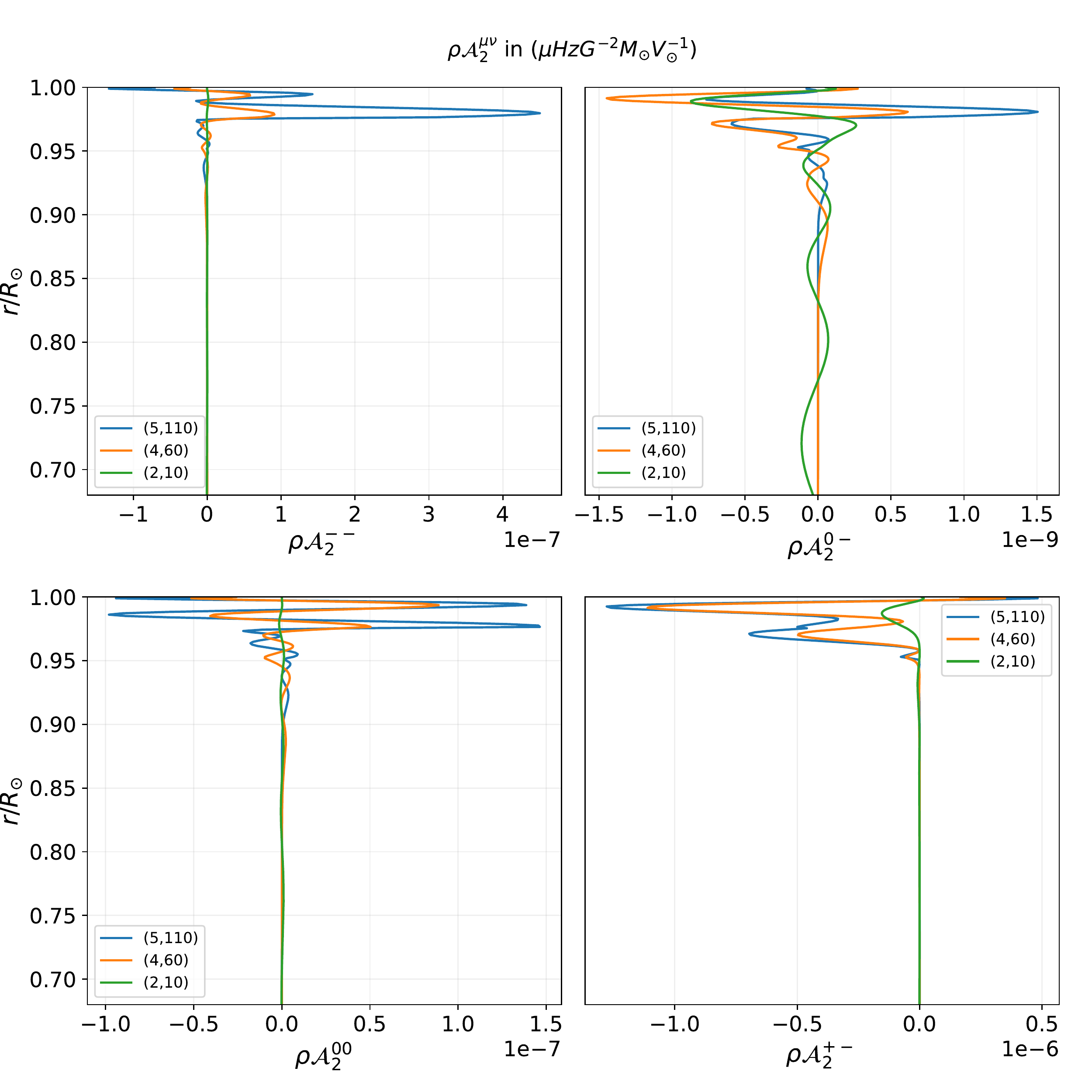} 
\caption{The four independent components of density-scaled sensitivity kernels $\rho\, \mathcal{A}_{2}^{\mu\nu}$ for $h_{20}^{\mu\nu}$. Kernels are plotted down to the base of the tachocline, $r \sim 0.68 R_{\odot}$. As shown in these plots, we have chosen three different $(n,\ell)$ for a  $t=0$ axisymmetric field. $\rho\, \mathcal{A}_{2}^{00}$ and $\rho\, \mathcal{A}_{2}^{+-}$ are sensitivities to isotropic components $h_{20}^{00}$ ($\propto \langle B_{r}B_{r} \rangle$) and $h_{20}^{+-}$ ($\propto \langle B_{\theta}B_{\theta} + B_{\phi}B_{\phi} \rangle$) respectively. $\rho\, \mathcal{A}_{2}^{0-}$ and $\rho\, \mathcal{A}_{2}^{--}$ are sensitivities to anisotropic components $h_{20}^{0-}$ (related to $\langle B_{r}B_{\theta} \text{ and } B_{r}B_{\phi} \rangle$) and $h_{20}^{--}$ (related to $\langle B_{\theta}B_{\phi} \rangle$) respectively.
} 
\label{fig: kernels_1D}
\end{figure}

\begin{figure}[t]
\centering
\includegraphics[width=1.03\linewidth]{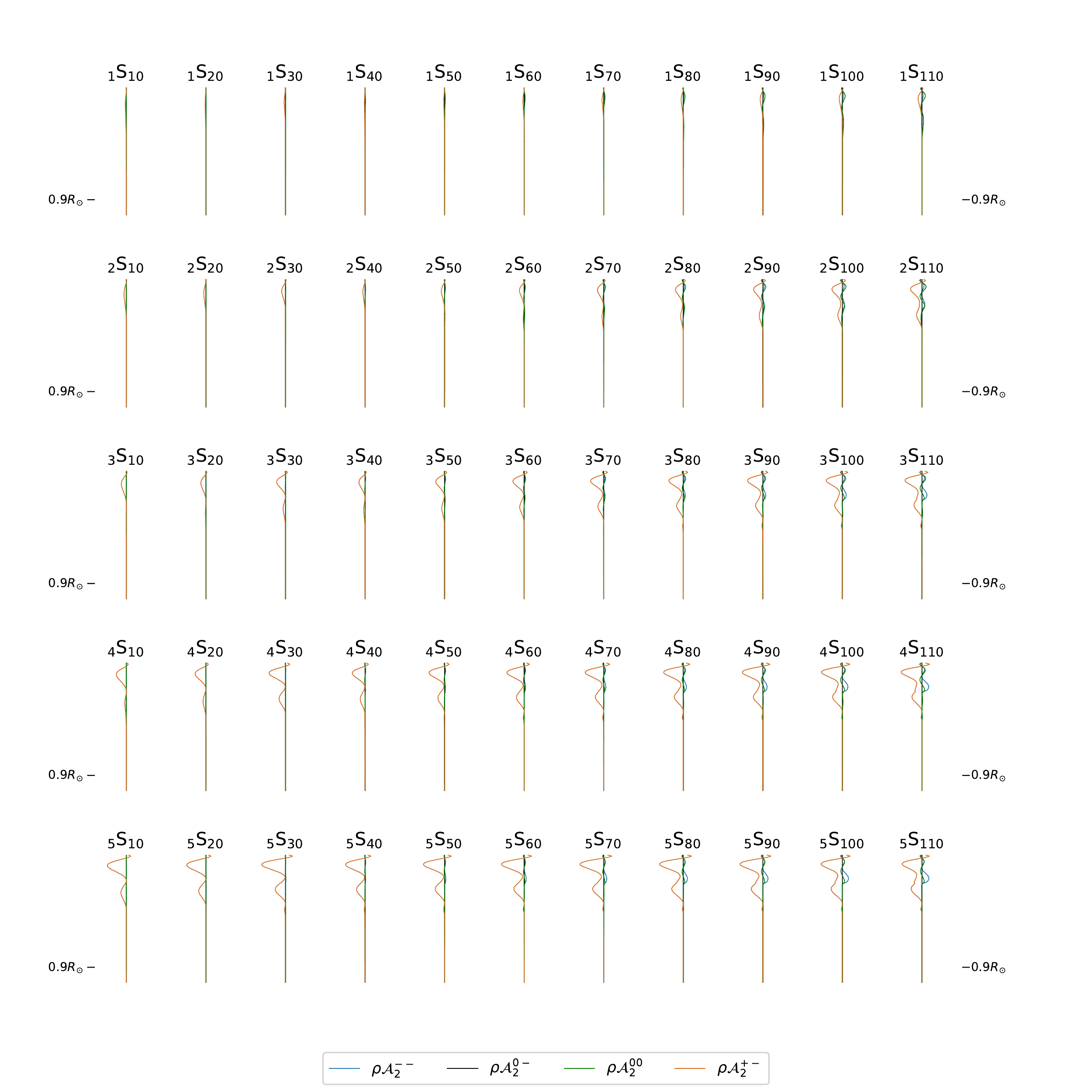}
\caption{Density-scaled sensitivity kernels $\rho\, \mathcal{A}_{2}^{\mu\nu}$ for Lorentz stress corresponding to self-coupling of spheroidal modes ${}_{n}\rm{S}_{\ell}$. As in Figure~\ref{fig: kernels_1D}, the $y$-axis is depth from the surface of the sun and is plotted up to $0.88 R_{\odot}$ and the $x$-axis is the strength of the kernel. We keep the $x$-range the same in all the plots for easy of comparison across varying $n,
\ell$. We have avoided marking the axes for the sake of clarity. However, to get a sense of the $x$-scale, one can compare the kernels for the very last mode ${}_{5}\rm{S}_{110}$ with those plotted in Figure~\ref{fig: kernels_1D}. We know that $\mathcal{A}_{s}^{\mu\nu} \propto \mathcal{B}_{s0}^{\mu\nu}$ (Eqns.~\ref{eqn:wigner_P_ljm_relation} and \ref{eqn:Kernel_m_separation}). Expressions for $\mathcal{B}_{s0}^{\mu\nu}$ may be found in Section~\ref{sec:mag_kern}. These kernels may be used to invert for $h_{20}^{\mu\nu}$ from measurements of $a_2^{n\ell}$.
}
\label{fig: a2_kern_allcompare}
\end{figure}

Figure~\ref{fig: kernels_1D} shows sensitivity kernels ($\rho\, \mathcal{A}_{s}^{\mu\nu}$) for the $a_{s}^{n\ell}$ coefficients due to density-scaled axisymmetric Lorentz-stress $h_{s0}^{\mu\nu}/\rho$ (and therefore effectively $h_{s0}^{\mu\nu}$), where $s=2$. We have chosen three distinct modes --- (5,110); (4,60); (2,10). The kernels have been plotted down to a depth of $r = 0.68 R_{\odot}$. This is because the equatorial location of the tachocline is $\sim0.7 R_{\odot}$ with a thickness of $\sim0.04 R_{\odot}$ \citep{howe09}. The sensitivity for the isotropic or on-diagonal component $h_{20}^{+-}$ is larger, by at least an order of magnitude, than all the other components. The other isotropic component $h_{20}^{00}$ and the anisotropic or off-diagonal components $h_{20}^{--}$ have comparable sensitivities. The anisotropic component $h_{s0}^{0-}$ has the least sensitivity. This is in agreement with \cite{hanasoge17} with an additional finding that $h_{20}^{--}$ has sensitivities comparable to $h_{20}^{00}$. This difference is because of the correction terms included in this study to the Lorentz-stress sensitivity kernels (Eqn.~\ref{eqn: cp_mat_corr}). Within the outer envelop $r \gtrsim 0.9 R_{\odot}$, the mode ${}_{5}\rm{S}_{110}$ is dominantly sensitive followed by ${}_{4}\rm{S}_{60}$, while ${}_{2}\rm{S}_{10}$ is the weakest of all components. This trend of sensitivity across $n,\ell$ is more evident in Figure~\ref{fig: a2_kern_allcompare}. Also, for the same kernel component, the mode possessing the maximum sensitivity changes with depth. For $h_{20}^{--}, {}_{2}\rm{S}_{10}$ becomes dominantly sensitive for $r < 0.9 R_{\odot}$. This holds true for the other components as well (see Figure~\ref{fig:110_dom_deeper}) where kernels have been plotted from 0.95 $R_{\odot}$ -- 0.68 $R_{\odot}$. This shows that there are broadly two category of modes --- surface and depth sensitive. 


Figure~\ref{fig: a2_kern_allcompare} shows the trend in variation of the sensitivity kernels $\rho\, \mathcal{A}_{2}^{\mu\nu}$ for self-coupling of spheroidal modes ${}_{n}\rm{S}_{\ell}$ for radial order $n = 1$--5 and every $10^{\rm{th}}$ angular degree $\ell = 10$--110. For a fixed radial order $n$, the near-surface sensitivity increases steadily with angular degree $\ell$. This increase in sensitivity is greatest for the kernel component $\mathcal{A}_{2}^{+-}$ which corresponds to the sensitivity for $\langle B_{\theta}B_{\theta} \rangle$ and $\langle B_{\phi}B_{\phi} \rangle$.
Although weaker than $\mathcal{A}_{2}^{+-}$, the kernel components $\mathcal{A}_{2}^{00}$ and $\mathcal{A}_{2}^{--}$ demonstrates a similar trend in sensitivity for increasing $\ell$. $\mathcal{A}_{2}^{00}$ corresponds to the sensitivity of the field component $\langle B_r B_r \rangle$ and $\mathcal{A}_{2}^{--}$ to $\langle B_{\theta}B_{\theta} \rangle$, $\langle B_{\phi} B_{\phi} \rangle$ and $\langle B_{\theta}B_{\phi} \rangle$. Therefore, the kernels have the largest sensitivity to the magnetic pressure and cross-terms of the angular components $\langle B_{\theta}B_{\phi} \rangle$ of Lorentz-stress. A faster rise in sensitivity is seen with increasing radial order $n$ for a fixed angular degree $\ell$. The cross-terms between radial and angular components, viz., $\langle B_r B_{\theta} \rangle$ and $\langle B_r B_{\phi} \rangle$ have poor sensitivities ($\mathcal{A}_{2}^{0-}$) and therefore more challenging to invert for. 

When inverting for an axisymmetric field,
Eqn.~(\ref{eqn:a_coeffs_invprob_expanded}) allows us to extract $h_{s0}^{00}$, $h_{s0}^{+-}$, $h_{s0}^{--}(-1)^{s} + h_{s0}^{--}$ and $h_{s0}^{0-}(-1)^{s} + h_{s0}^{0+}$. Here, $h_{s0}^{00}$ and $ h_{s0}^{+-}$ are the isotropic or diagonal components and $h_{s0}^{--}(-1)^{s} + h_{s0}^{++}$ and $ h_{s0}^{0-}(-1)^{s} + h_{s0}^{0+}$ are the anisotropic or off-diagonal components of $\boldsymbol{\mathcal{H}}$. Using relations~(\ref{eqn: BrBr_App})--(\ref{eqn: BphiBphi_App}) along with the identity $Y_{l0}^{-N} = (-1)^{N} Y_{l0}^N$ \citep[see Appendix C of][]{DT98}, the connection between axisymmetric Lorentz-stress components in the GSH basis and the ($r,\theta,\phi$) basis is as follows:
\begin{eqnarray}
    B_r B_r &=& \sum_s h_{s0}^{00}\,Y_{s0}^{0}, \label{eqn: BrBr}\\
    B_{\theta} B_{\theta} + B_{\phi} B_{\phi} &=& -\sum_{s} 2h_{s0}^{+-} \,Y_{s0}^0, \label{eqn: BtBt_BpBp}\\
    B_{\theta} B_{\theta} - B_{\phi}B_{\phi} &=& \sum_{s} (h_{s0}^{++}\,Y_{s0}^{+2} + h_{s0}^{--}\,Y_{s0}^{-2}) = \sum_s (h_{s0}^{++} + h_{s0}^{--})\,Y_{s0}^{-2}, \label{eqn: BtBt_BpBp2}\\
    B_r B_{\theta} &=& \tfrac{1}{\sqrt{2}}\,\sum_s  (h_{s0}^{0-}\,Y_{s0}^{-} - h_{s0}^{0+}\,Y_{s0}^{+}) = \tfrac{1}{\sqrt{2}}\,\sum_s (h_{s0}^{0-} + h_{s0}^{0+})\,Y_{s0}^{-}, \\
    B_r B_{\phi} &=& - \tfrac{i}{\sqrt{2}}\, \sum_{s} (h_{s0}^{0-}\,Y_{s0}^{-} + h_{s0}^{0+})\,Y_{s0}^{+} = - \tfrac{i}{\sqrt{2}}\, \sum_{s} (h_{s0}^{0-} - h_{s0}^{0+})\,Y_{s0}^{-}, \\
    B_{\theta} B_{\phi} &=& \tfrac{i}{2}\, \sum_{s} (h_{s0}^{++}\,Y_{s0}^{+2} - h_{s0}^{--}\,Y_{s0}^{-2}) = \tfrac{i}{2}\, \sum_{s} (h_{s0}^{++} - h_{s0}^{--})\,Y_{s0}^{-2} \label{eqn: BtBp}.
\end{eqnarray}
The implications of the above set of relations and the expressions for the kernels listed in Appendix~\ref{sec:mag_kern} may be summarized as follows.
\begin{enumerate}
    \item The isotropic components may be directly extracted from an inverse problem (Eq.~[\ref{eqn:a_coeffs_invprob_expanded}]). Equations~(\ref{eqn: BrBr}) and~(\ref{eqn: BtBt_BpBp}) relate these components to $\langle B_r B_r\rangle,\langle  B_{\theta}B_{\theta} + B_{\phi}B_{\phi}\rangle$ and therefore to magnetic pressure $B^2/8\pi$. However, on account of the $(1+p)$ factor in $\cB_{s0}^{00}$ and $\cB_{s0}^{+-}$ in Eqns.~(\ref{eq:kern_00}) and~(\ref{eq:kern_pm}), where $p = (-1)^{s}$ (or $(-1)^{\ell+\ell'+s}$ for cross-coupling), only even-$s$ components may be inferred. Because of the absence of odd-$s$ in the isotropic components, self coupling renders the inference of $\langle B_r B_r \rangle$, $\langle B_{\theta}B_{\theta} + B_{\phi}B_{\phi}\rangle$ incomplete, as according to Eqns.~(\ref{eqn: BrBr}) and~(\ref{eqn: BtBt_BpBp}). This may be bypassed when working with cross coupling of modes where $\ell+\ell'$ is not necessarily even, e.g., in normal-mode coupling studies.
    \item Eqn.~(\ref{eqn:a_coeffs_invprob_expanded}) allows the inference of a linear combination of anisotropic terms. For self coupling, these components are related to $\langle B_r B_{\phi} \rangle, \langle B_{\theta}B_{\phi} \rangle$ for odd $s$ and $\langle B_r B_{\theta} \rangle, \langle B_{\theta}B_{\theta} - B_{\phi}B_{\phi} \rangle$ for even $s$. This is because odd $s$ give the components of $(h_{s0}^{0-} - h_{s0}^{0+})$ and $(h_{s0}^{++} - h_{s0}^{--})$ while even $s$ give the components of $(h_{s0}^{0-} + h_{s0}^{0+})$ and $(h_{s0}^{++} + h_{s0}^{--})$. Kernels $\cB_{s0}^{0-}, \cB_{s0}^{--}$ are sensitive to both even and odd $s$. Once again, for self coupling, the complete inference of Lorentz-stress components in physical space is hindered because of the absence of either even- or odd-$s$ terms in the summations in Eqns.~(\ref{eqn: BtBt_BpBp2})-(\ref{eqn: BtBp}). 
    \item For both the cases above, limitations in inverting for components of $\Bv\Bv$ may be overcome in studies where coupling across multiplets is considered. For example, $(h_{s0}^{0-} + h_{s0}^{0+})$ and $(h_{s0}^{++} + h_{s0}^{--})$ for $s$ being odd (even) may be recovered by choosing $(\ell'+\ell)$ odd (even). This gives us access to $h_{s0}^{0-} \pm h_{s0}^{0+}$ and $h_{s0}^{++} \pm h_{s0}^{--}$ for all $s>1$ and therefore all components of $h_{s0}^{\mu\nu}$ for $s>1$. Therefore, the summation over $s$ in Eqns.~(\ref{eqn: BrBr})-(\ref{eqn: BtBp}) would be complete (excluding $h_{10}^{--}$ or equivalently $h_{10}^{++}$).
    \item These inversions cannot recover components $h_{00}^{--}(-1)^{s} + h_{00}^{++}$, $h_{00}^{0-}(-1)^{s} + h_{00}^{0+}$ and  $h_{10}^{--}(-1)^{s} + h_{10}^{++}$ as they correspond to kernels proportional to $\mathcal{B}_{00}^{--}, \mathcal{B}_{00}^{0-}$ and $\mathcal{B}_{10}^{--}$. These kernels are proportional to $\wigfull{\ell}{s}{\ell}{-p}{q}{p}$, which is zero for $s < |q|$ (see Section~\ref{sec:mag_kern}). We consider only deviations from spherical symmetry here, i.e., $s\ge1$.  
    Therefore, for the $\ev{+}\ev{+}$ or $\ev{-}\ev{-}$  component the $s=1$ angular degree is not accessible through inversions. This is true even for cross-modal coupling. 
\end{enumerate}

\subsection{Modeling a custom magnetic field at solar minimum} \label{sec: Theory3}

\begin{figure}[ht]
\centering
\includegraphics[width=0.8\linewidth]{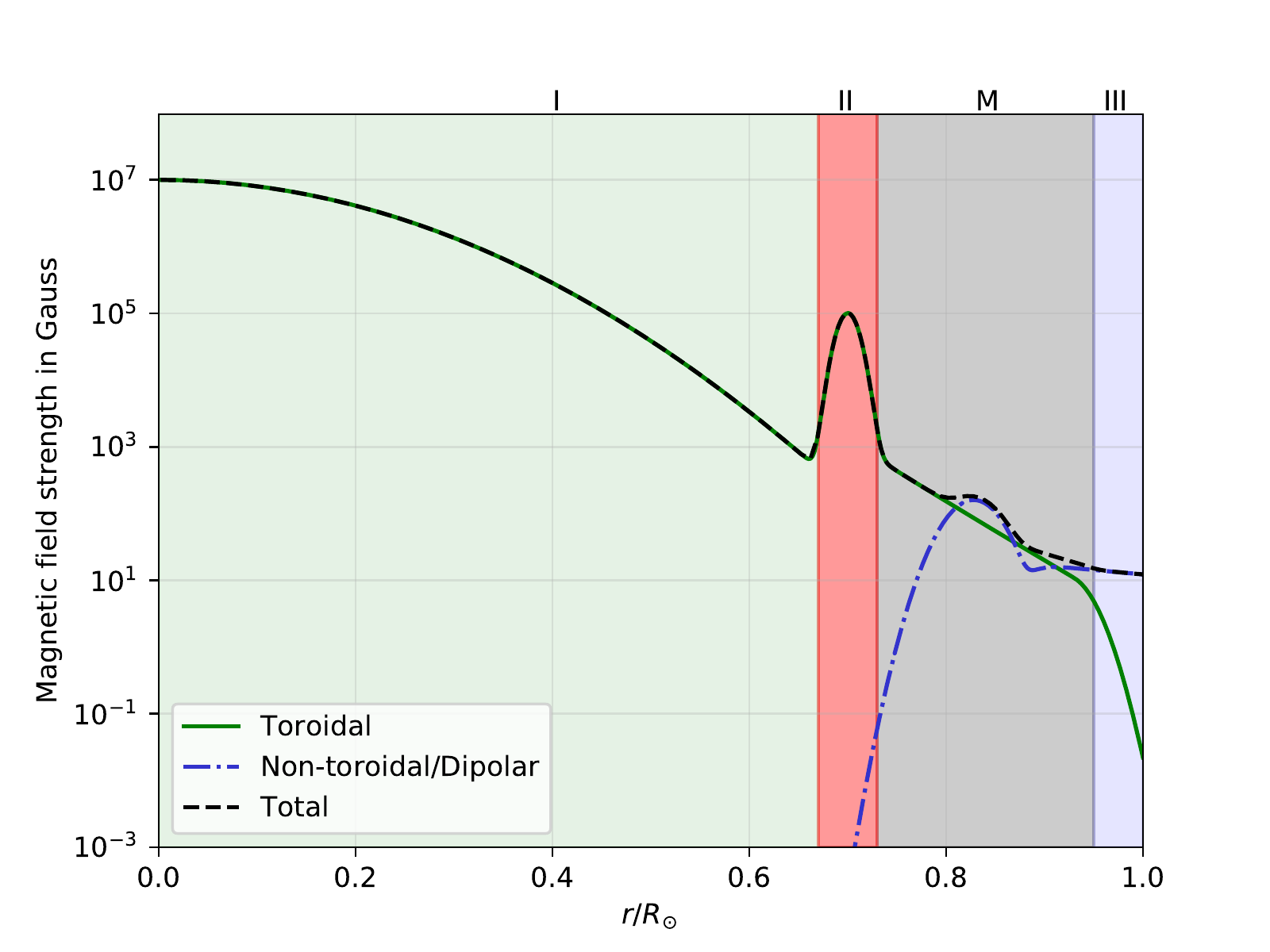}
\caption{Magnetic field strength associated with three different regions with the following field configurations: (A) Purely toroidal for $r<0.72 R_{\odot}$ containing the tachocline at $r = 0.7 R_{\odot}$, (B) Toroidal and spheroidal extending from $r = 0.72 R_{\odot}$ to $r = 0.95 R_{\odot}$ (marked by the \textit{grey} shaded region) and (C) Purely dipolar for $r > 0.95 R_{\odot}$. The expression for $\Bv$ is found in Eqn.~(\ref{eqn: mag_field}) and the plot of the associated parameters is shown in Figure~\ref{fig:field_params}.} \label{fig: field_strength_plot}
\end{figure}

The primordial magnetic field trapped in the core is believed to be dominantly toroidal \citep{GoughTaylor1984,Parker55} and so is the field at the tachocline \citep{Antia2000,nandy02}. Nevertheless, during the solar minimum, the surface magnetic field is roughly dipolar \citep{Andres13,Bhowmik2018}. In trying to analytically model the simplest version of a similar global magnetic field, we construct the following
\begin{enumerate}
    \item Region I: Extending from the center of the Sun to the inner base of the tachocline at $r=0.68 R_{\odot}$. The field is purely toroidal and its strength falls off radially as a Gaussian peaked at $r=0$. We refer to the magnetic field strength around $r=0$ as $B_I$. 
    \item Region II: Extending over the shell $0.68 R_{\odot} < r < 0.72 R_{\odot}$ this region marks the tachocline. Here again, the field has a purely toroidal configuration. The magnetic strength peaks at $r=0.7R_{\odot}$. We call this field strength $B_\text{II}$.
    \item Region M: Extending over the shell $0.72 R_{\odot} < r < 0.95 R_{\odot}$, this region is marked by a `mixed' field that allows a smooth transition from an inner toroidal to an outer dipolar field. The strength of the toroidal field dominates in $0.72 R_{\odot}<r<0.81R_{\odot}$ and in $0.86 R_{\odot}<r<0.91R_{\odot}$ while the spheroidal component dominates the other parts. However, both components contribute non-trivially in this region.
    \item Region III: Extends outward from $r=0.95 R_{\odot}$ up to the solar surface. This is purely dipolar and has a strength of $B_\text{III}$ at the surface. For the Sun, $B_\text{III}\sim 10G$. This allows the field to smoothly match the photospheric dipolar configuration (found at solar minima).
\end{enumerate}\
Figure~\ref{fig: field_strength_plot} shows the radial variation of field strength according to the regions illustrated above. The green line shows the strength of the toroidal field as a function of radius. The blue line shows the strength of (a) the spheroidal field in regions --- II and M, and (b) the purely dipolar field in region III. The extremely low magnitude of the toroidal field (spheroidal field) in Region III (Region I \& II) indicates that Region III (Region I \& II) is almost completely dipolar (toroidal). Further details about the construction of the field strength and configuration may be found in Section~\ref{sec:B_construction} and Figure~\ref{fig:field_params}. For the purpose of the forward problem, we explore a set of eight models (Table~\ref{tab:model_list}) with varying relative strengths of $B_\text{I}, B_\text{II}$ and $B_\text{III}$. Each of these can take ``Low" and ``High" values. The ``High" field strength exceeds the ``Low" field strength by a factor of 10. Model M1, where all three regions have a ``Low" field, qualitatively resembles the Sun-like case, where $B_\text{I} = 10^7 G, B_\text{II} = 10^5 G$ and $B_\text{III} = 10 G$.

\begin{table}[ht]
\centering
\begin{tabular}{|c|c|c|c|c|c|c|c|c|}
\hline
\textbf{} & M1 & M2 & M3 & M4 & M5 & M6 & M7 & M8 \\ \hline
$B_\text{I}$      & Low         & Low         & Low         & Low         & High        & High        & High        & High        \\ \hline
$B_\text{II}$      & Low         & Low         & High        & High        & Low         & Low         & High        & High        \\ \hline
$B_\text{III}$      & Low         & High        & Low         & High        & Low         & High        & Low         & High        \\ \hline
\end{tabular}
\caption{Table describing the different cases of model magnetic fields we use for our forward problem. $B_\text{I}, B_\text{II}, B_\text{III}$ are the peak magnitudes of magnetic fields in Region I (at the center, $r = 0$), Region II (at the tachocline, $r = 0.7 R_{\odot}$) and Region III (at the surface $r = R_{\odot}$), respectively. Fields marked ``High" are stronger than fields marked ``Low" by a factor of 10.}
\label{tab:model_list}
\end{table}

\begin{table}[ht]
\centering
\begin{tabular}{|c|c|c|c|c|}
\hline
Modes          & \multicolumn{2}{c|}{${}_{2}\rm{S}_{10}$}   & \multicolumn{2}{c|}{${}_{5}\rm{S}_{110}$}  \\ \hline
$a$-coefficients & \textbf{$a_0$} & \textbf{$a_2$} & \textbf{$a_0$} & \textbf{$a_2$} \\ \hline
M1    & 4.138         & -0.501        & 2.399         & 0.628        \\ \hline
M2    & 4.991         & -0.167        & 239.986       & 62.863       \\ \hline
M3    & 399.696       & -62.360       & 2.399         & 0.628        \\ \hline
M4    & 400.549       & -62.026       & 239.986       & 62.863      \\ \hline
M5    & 14.274        & 11.824        & 2.399         & 0.628        \\ \hline
M6    & 15.127        & 12.158        & 239.986       & 62.863       \\ \hline
M7    & 411.551       & -50.349      & 2.399         & 0.628        \\ \hline
M8    & 412.405       & -50.015      & 239.986       & 62.863      \\ \hline
\end{tabular}
\caption{List of $a_0^{n\ell}$ and $a_2^{n\ell}$ for modes $(n,\ell) = {}_{2}\rm{S}_{10}$ and ${}_{5}\rm{S}_{110}$. All values are in nHz. M1--M8 indicate the models described in Table \ref{tab:model_list}. These models vary in magnetic strengths in the core (Region I), the tachocline (Region II), and the surface (Region III).}
\label{tab:a_coeffs}
\end{table}

We compute $a$-coefficients for two modes ${}_{2}\rm{S}_{10}$ and ${}_{5}\rm{S}_{110}$ (plotted in Figure~\ref{fig: kernels_1D}), which possess different regions of sensitivity. ${}_{5}\rm{S}_{110}$ has a shallow but strong sensitivity to Lorentz stresses while ${}_{2}\rm{S}_{10}$ has a comparatively weak but radially extended sensitivity (Figure~\ref{fig:110_dom_deeper}). Therefore, the former may be used to study near-surface fields with high precision while the latter may be used to probe much deeper fields. This is evident from Table \ref{tab:a_coeffs} where ${}_{5}\rm{S}_{110}$ has the same value of $a$-coefficients up to a precision of atleast 1pHz for models (M1, M3, M5, M7) with a ``low" surface field (irrespective of the field strength in the interior regions I and II). Similarly, the models (M2, M4, M6, M8) with a ``high" surface field have the same $a$-coefficients, and possess negligible sensitivity to the interior field strength variation. On fixing surface field strength ($B_\text{III}$) but varying interior field strengths ($B_\text{I}$ and $B_\text{II}$) for the mode ${}_{5}\rm{S}_{110}$, we found the value of $a$-coefficients to vary on the order of 1e-9 nHz, which is well below the present limit of resolution. 

In the Sun-like model M1, for ${}_{2}\rm{S}_{10}$ the mean shift of frequency, $a_0^{n,\ell}$, is larger than that for ${}_{5}\rm{S}_{110}$. This is on account of the fact that large values of magnetic fields near the tachocline (and inwards towards the core) are also accounted for by the deeper sensitivities of $\mathcal{A}_{10}^{\mu\nu}$. We find a weak increase in the magnitude of $a_0^{2,10}$ from M1 to M2 (or equivalently from M3 to M4). In contrast, there is an increase by a factor of 100 in $a_0^{5,110}, a_2^{5,110}$. This highlights the greater near-surface sensitivity of ${}_{5}\rm{S}_{110}$. In going progressively from M1 to M4, the magnetic field strength near the tachocline ($B_\text{II}$) and the surface ($B_\text{III}$) increases while the core field ($B_\text{I}$) stays constant. The magnitude of $a_0^{2,10}$ and $a_2^{2,10}$ increases from M1 (or M2) to M3 (or M4) approximately by a factor of 100. This shows the extremely high sensitivity of ${}_{2}\rm{S}_{10}$ to the field near the tachocline (Region II). The comparatively weaker, yet still noticeable, increase from M3 to M4, once again, shows the weak sensitivity of ${}_{2}\rm{S}_{10}$ to magnetic fields near the surface. A similar analysis may be performed on the magnitudes of $a_0^{2,10}, a_2^{2,10}$ when going progressively from M5 to M8. Finally, comparing $a$-coefficients of M1 with M5 (or M[i] with M[i+4] where i = 1,2,3,4), there is a mean increase of approximately 10 nHz. This marks the non-negligible sensitivity of ${}_{2}\rm{S}_{10}$ to the presence of fields beyond the tachocline towards the stellar interior. This is roughly in keeping with \cite{Kiefer18} where they note that the effect of magnetic fields in the radiative interior can only be of the order of nHz.

The inclusion of differential rotation with an (mis)aligned axis of symmetry with the magnetic axis will be the focus of a future study. A misaligned magnetic and rotation axis would necessitate the rotation of GSH \citep{Edmonds} and therefore entail significant mathematical complexity and computational burden. 

\subsection{Splitting functions} \label{sec: splitting_functions_results}

In this section we compute splitting functions due to self-coupled and cross-coupled multiplets \citep[detailed treatment may be found in Chapter 14 of][]{DT98}.  Constructing the map of a splitting function $\eta^{n'\ell' n\ell} = \sum_{s,t} c_{st}^{n'\ell' n\ell}\,Y_{st}$ demands knowledge of the structure coefficients $c_{st}^{n'\ell' n\ell}$ (see Eqn~[\ref{eqn: inv_prob_cst}]).
Expressions for $\cG_{s}^{\mu\nu}$ are given in Eqns.~$(\ref{eq:kern_mm})-(\ref{eq:kern_pm})$. Therefore, for the forward problem, we need to construct $h_{st}^{\mu\nu}$ or $\boldsymbol{\cH}$ to obtain the splitting functions. We do so by choosing a $\Bv$ field which is real and solenoidal ($\bnabla\cdot\Bv = 0$). The realness of $\Bv$ implies $(B_{st}^{\mu})^* = (-1)^t B_{s\bar{t}}^{\bar{\mu}}$ and hence also $(h_{st}^{\mu\nu})^* = (-1)^t h_{s\bar{t}}^{\bar{\mu}\bar{\nu}}$. As mentioned earlier, $\bar{\mu} = -\mu$ and so on. Further, solenoidality imposes the constraint $B_{st}^{+} + B_{st}^{-} = \partial_r (r^2 B_{st}^0)/r\Omega_s^0$. We choose the GSH components of $\Bv$ as follows:
\begin{eqnarray}
    B_{st}^0(r) &=& \mathcal{R}(s,t) b_0(r), \label{eqn:B_st^0_random}\\
    B_{st}^+(r) &=& \mathcal{R}(s,t) \partial_r(r^2 B_{st}^0)/r\Omega_s^0, \\
    B_{st}^-(r) &=& \partial_r(r^2 B_{st}^0)/r\Omega_s^0 - B_{st}^+(r), \label{eqn:B_st^-_random}
\end{eqnarray}
where $\mathcal{R}(s,t)$ is a set of random numbers between 0 and 1, so as to construct a generic field configuration. The total field strength shown in Figure~\ref{fig: field_strength_plot} is $b_0(r)$. We then impose the constraint $(B_{st}^{\mu})^* = (-1)^t B_{s\bar{t}}^{\bar{\mu}}$. Thereafter we construct $\cH$ according to Eqn.~(\ref{eqn: generic_Hmunu}). At this point we have the coefficients~$c_{st}^{n'\ell' n\ell}$ and can plot the splitting function $\eta^{n'\ell' n\ell}$. It should be noted that we needed to construct artificial structure coefficients as this study does not harness actual solar data. When using real observations, $c_{st}^{n'\ell'n\ell}$ ought to be inferred from inversions. 
\begin{figure}[h]
\begin{minipage}{0.47\textwidth}
    \includegraphics[width=\linewidth]{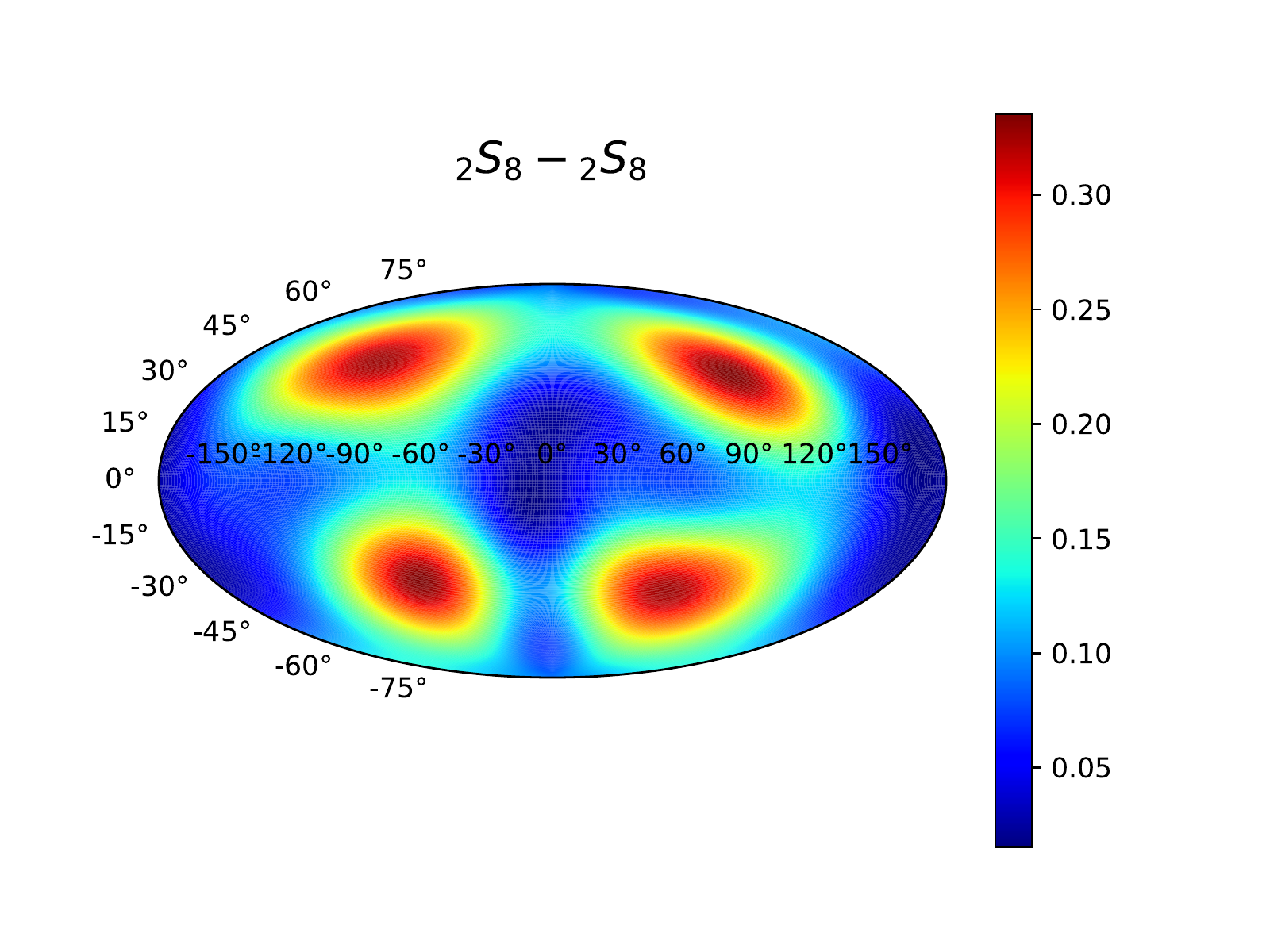}
    \end{minipage}
    \hspace{\fill} 
    \begin{minipage}{0.47\textwidth}
    \includegraphics[width=\linewidth]{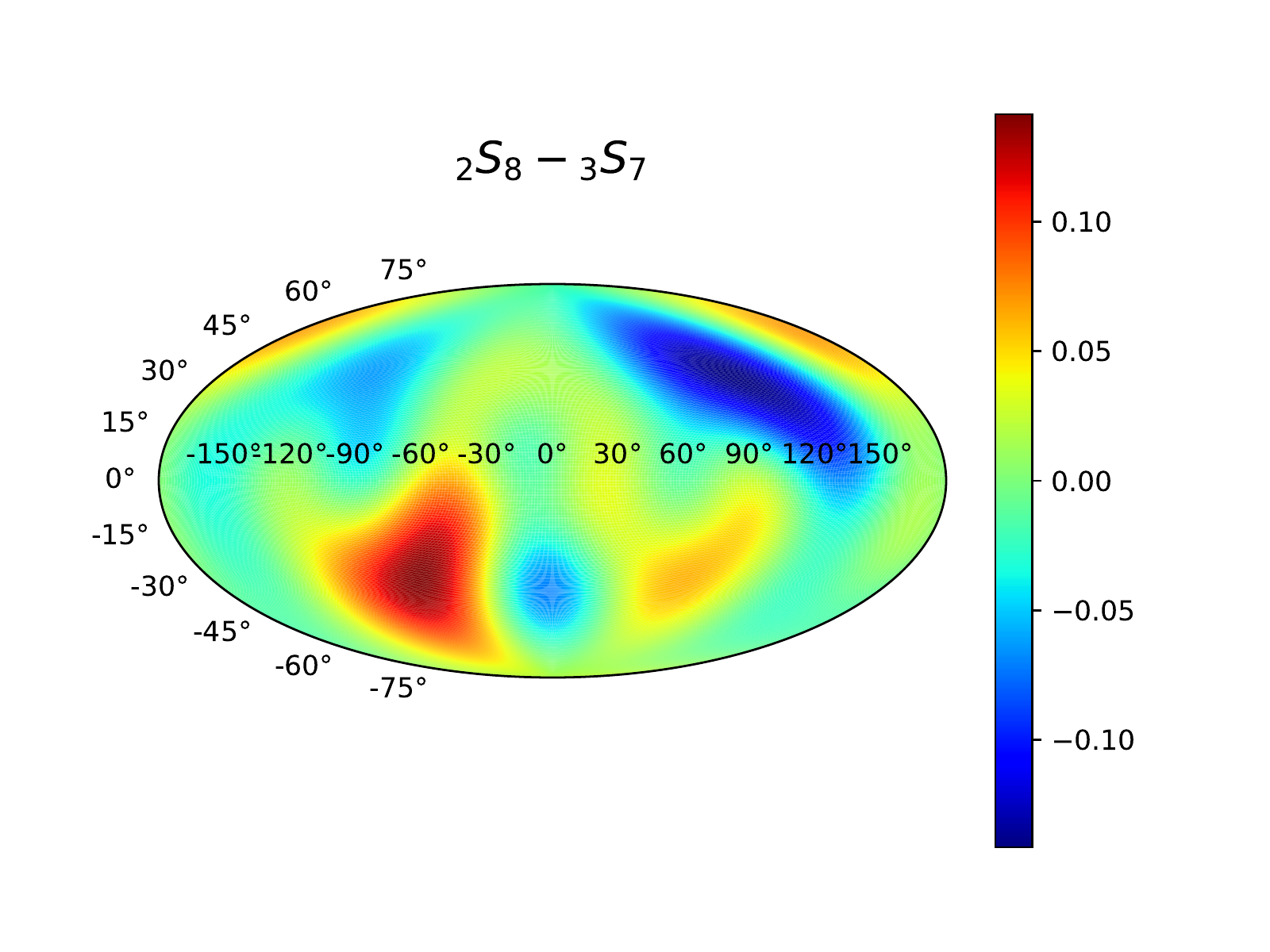}
    \end{minipage}
\caption{``Aitoff" projection of splitting functions $\eta(\theta,\phi)$ for self-coupling of multiplet $\mode{2}{8}$ and cross-coupling of $\mode{2}{8}$ with $\mode{3}{7}$. We construct a real, solenoidal artificial $\Bv$ field. The strength of the various components $B_{st}^{\mu}(r)$ were scaled with random numbers (as shown in Eqns.~[\ref{eqn:B_st^0_random}]~$-$ß~[\ref{eqn:B_st^-_random}]) over $b_0(r)$ which is the 'total' field strength shown in Figure~\ref{fig: field_strength_plot}. The artificial $B^{\mu}_{st}$ profiles have $s=1,2,3$ and therefore the artificial $h_{st}^{\mu\nu}$ have $s=0,1,2,3,4,5,6$. The colorbar is dimensionless and scales as the fractional change in frequency. The values are not supposed to be a representation of the actual Sun and are certainly an over-estimation of the true solar splitting functions under Lorentz-stress perturbations.} \label{fig:splitting_functions}
\end{figure}

We plot the splitting functions in Figure~\ref{fig:splitting_functions} for two cases: (a) self-coupling of $\mode{2}{8}$, and (b) cross-coupling of $\mode{2}{8}$ with $\mode{3}{7}$. The unperturbed frequencies ${}_2\omega{}_{8}$ and ${}_3\omega{}_{7}$ are less than $5 \mu$~Hz apart. The colormaps have been plotted at the solar surface $r=R_{\odot}$. This means that any point in $(\theta,\phi)$ reflects the structure due to Lorentz-stress field $h_{st}^{\mu\nu}$ integrated along the radial direction, weighted by the respective kernel components $\cG_{s}^{\mu\nu}$ (ref. Eqn~[\ref{eqn: inv_prob_cst}]). Thus, as mentioned earlier in Section~\ref{sec: coupled_multiplets}, the splitting functions are a convenient way to visualize the internal structure perturbations as seen ``through" its kernels. A zero value of splitting functions at a certain $(\theta_0,\phi_0)$ does not necessarily mean $h_{st}^{\mu\nu}(r,\theta_0,\phi_0) = 0$, i.e, it need not be inferred that there is no structure beneath that surface location. It could simply means that either the kernel sensitivity is negligible where the structure exists or that the kernel-weighted structure averages out to zero.

\section{Discussions and Conclusions}\label{sec: discussion&conclusion}

In this study, we have addressed the long-standing problem of proposing a formalism to calculate the effect of a completely general Lorentz stress field (Eqn~[\ref{eqn: H_exp}]) on shifts in solar eigenfrequencies (Eqn~[\ref{eqn:evalue_problem_QDPT}]). We initially adopted the approach of quasi-degenerate perturbation theory, thereby allowing for cross coupling between modes which are sufficiently close in frequency. The coupling matrix $\Lambda_{k'k}$ for full coupling may be found in Eqn.~(\ref{eqn: lambda_m}). As shown in Section~\ref{sec: isolated_multiplet}, the theory may be readily reduced to a degenerate perturbation analysis (and hence the study of isolated-multiplets) in a straightforward manner. However, the isolated-multiplet assumption hinges on the existence of multiplets well-separated in frequency from any other multiplet, compared to the frequency shifts induced due to the perturbing force. We first demonstrate the existence of such isolated-multiplets for the case of Lorentz-stresses in the Sun (refer to Section~\ref{sec: mode_labelling}). For the sake of completeness,
we have also demonstrated that frequency shifts due to differential rotation can be explained by considering self-coupling only (see Appendix~\ref{sec: DR_mode_labelling}). The novelty of our work is in proposing and analyzing the kernels for (a) general magnetic fields and (b) the specific case of dealing with isolated multiplets under an axisymmetric magnetic field --- the $a$-coefficient formalism. 

We formulated the forward problem by respecting cross-mode coupling and a general field configuration. In doing so, we use the idea of so-called ``structure coefficients" in Section~\ref{sec: coupled_multiplets}. For the purpose of illustration, we use a real and solenoidal magnetic field possessing angular degrees $s=1,2,3$. Borrowing from geophysical literature, we visualize the effect of frequency splitting under self- and cross-coupling of modes via splitting functions in Section~\ref{sec: splitting_functions_results}. We then characterize the impact of simple analytically modelled axisymmetric magnetic-fields on frequency shifts in an isolated multiplet, method of $a$ coefficients (see Sections~\ref{sec: a-coefficients} and \ref{sec: Theory3}). These models share the same geometric configuration (inner toroidal, outer near-surface dipolar and an intermediate mixed field). Varying peak magnetic-field strengths in these three regions by a factor of 10 results in a total of eight models (Table~\ref{tab:model_list}). $a$-coefficients obtained from these models (Table~\ref{tab:a_coeffs}) demonstrate the significant difference in sensitivities of different modes to different regions in the Sun. We choose two modes --- ${}_{5}\rm{S}_{110}$ sensitive to shallow layers and ${}_{2}\rm{S}_{10}$ sensitive to relatively deeper layers (see Figure~\ref{fig:110_dom_deeper}) --- to bring out the effect of changing the magnitude of the Lorentz stress at varying depths in the Sun. 

When analysing self-coupling kernels for $a$ coefficients, we find that there is a well-defined trend in the near-surface sensitivity when moving across $n$ and $\ell$ space (see Figure~\ref{fig: a2_kern_allcompare}). The sensitivity increases while moving to larger values of $n$ and $\ell$. Sensitivities of kernels near the surface are higher by at least an order in magnitude than in the interior. This would facilitate precise inference about near-surface Lorentz stresses and should assist in the validation of earlier observational or theoretical claims \citep{Antia2000,Lites2008,Petrie2009,baldner10}.
As also noted by \cite{Kiefer18}, the low angular degrees $\ell$ modes are sensitive to deeper layers (shown for a single case in Figure~\ref{fig:110_dom_deeper}). The model calculation for constructing an $s=1$ solenoidal field (outlined in Appendix~\ref{sec:B_construction}) is easily extended to the purpose of constructing general axisymmetric fields of arbitrary angular degree $s$. This may be essential for constructing artificial $a$-coefficient profiles for comparing with real observations in order to place constraints on field geometry and strength. Most previous studies \citep[e.g.,][]{goughmag,Antia2000,goode04,Kiefer17,Kiefer18} used a toroidal model magnetic field close to the surface. This is motivated by work on dynamo simulations \citep[e.g.,][]{Miesch16} who find toroidal field strengths dominating over poloidal. We choose a mixed field (dominantly toroidal) up to $r \leq 0.95 R_{\odot}$ followed by a purely dipolar field up to the surface so as to match the boundary with the global dipolar field observed during a solar minima.  For our model the boundaries of toroidal, mixed (both toroidal and spheroidal) and dipolar configurations can be easily tweaked to accommodate further complicated field geometries. This allows for a more realistic field geometry as it would comprise a combination of toroidal and spheroidal fields as opposed to purely toroidal fields \citep{Parker55}.

We discuss selection rules that govern general Lorentz-stress-induced mode coupling (Appendix~\ref{sec:mag_kern}) and within an isolated multiplet for an axisymmetric Lorentz stress in Section~\ref{sec: self-coupling-axisymmetric}. The inverse problem naturally dictates the recovery of the untangled-isotropic ($h_{st}^{00}, h_{st}^{+-}$) and the tangled-anisotropic ($h_{st}^{0-}(-1)^{\ell'+\ell+s} + h_{st}^{0+}, h_{st}^{--}(-1)^{\ell'+\ell+s}+h_{st}^{++}$) components of Lorentz stresses. This study identifies fundamental advantages of approaches that harness cross-mode couplings over the isolated-multiplet assumption to recover the full spectrum in angular degree of the Lorentz stresses. 

For studies using quasi-degenerate perturbation theory, the supermatrix $\mathcal{Z}_{k'k}$ which may be constructed from our prescribed kernels for a general real Lorentz stress field $\boldsymbol{\mathcal{H}}$ is Hermitian (due to the symmetry relation mentioned in property 5 in Appendix~\ref{sec:mag_kern} along with realness of $\boldsymbol{\mathcal{H}}$). This ensures real-valued frequencies and thereby stability of the perturbed modes.
This is contrary to the discussion in \citep{Kiefer17},
where they retain the possibility of having a non-hermitian $\mathcal{Z}_{k'k}$ in the context of purely toroidal fields. It is to be noted that they adopt a different way of expressing the coupling matrix which they call $H_{k'k}$ \citep[see Eqn.~(31) in][]{Kiefer17}. Their $H_{k'k}$ is expressed as a sum of 25 terms weighted by ``angular kernels" for toroidal magnetic fields and does not contain surface terms. The essential difference with our coupling matrix $\Lambda_{k'k}$ (as shown in Eqn.~[\ref{eqn: cp_mat_corr}]) arises because we isolate the Lorentz stress tensor $\cH$ and in doing so, end up with a mode-symmetric kernel for $\cH$ as well as a collection of surface terms as in Eqn.~(\ref{eq: mag_surface_terms}).

The $s=1$ mode is found to suffer from the limitation of being invertible for all but one independent component ($h_{1t}^{--}$) in the GSH basis. This is because the corresponding kernel $\cB_{1t}^{--}$ vanishes because of the selection rule imposed by the Wigner-3j symbols. As a result the coupling matrix elements are not sensitive to existence of $h_{1t}^{--}$. For an axisymmetric field, this implies that the $s=1$ component of $(\langle B_{\theta}B_{\theta} \rangle - \langle B_{\phi}B_{\phi} \rangle)$ and $\langle B_{\theta} B_{\phi} \rangle$ cannot be recovered using this formalism (see Eqns.~\ref{eqn: BtBt_BpBp2}, \ref{eqn: BtBp}, and ~\ref{eq:kern_mm}). Presently, translating the solenoidal constraint of the magnetic field to an equivalent condition in the Lorentz-stress components seems unlikely. However, if found, such a constraint could help recover or constrain these missing components. 

Our formalism is readily extended to asteroseismic cases, where modes up to $\ell=3$ are reliably observed \citep{Chaplin13}, although whether it is possible to extract new and useful information is unclear. This study adds to the list of earlier efforts directed toward constraining the interior magnetic field of the Sun and goes beyond to seek a formal inverse problem for general Lorentz stresses. Most of these studies, as mentioned earlier  \citep[with the exception of][]{baldner10}, were restricted to a toroidal field geometry. \cite{baldner10} found a combination of a poloidal field and a double-peaked near-surface toroidal field to best reproduce the observed $a$ coefficients. However, it suffered from the limitation of excluding cross terms in toroidal and poloidal fields in regions of overlap in the solar interior. This is because magnetic-field forcing comes as a non-linear term and therefore the effect of two different field configurations cannot be added. Our formalism is not limited by any of these constraints. It, therefore, allows a neat generation of frequency shifts or $a$ coefficients due to axisymmetric magnetic fields, rigorously accounting for all tensorial components of Lorentz stress. \cite{baldner10} also notes that the possibility of radial profiles of the toroidal fields is ``limitless", implying that the best-fit model they obtained through a forward problem need not be the only model that produces similar $a$ coefficients. In that regard, the approach of an inverse problem may be an attractive avenue.

It should be noted that, in principle, the presence of a second-rank Lorentz stress tensor should introduce toroidal components in the eigenfunctions of the background model to an otherwise spheroidal system. However, the observed solar spectrum is known to be very well explained by a solely spheroidal theory. Therefore, we believe that inaccuracies on account of ignoring the magnetically induced toroidal modes should be negligible. A possible way to include toroidal modes in the calculation would be to compute the next order in the expansion. This would involve re-evaluating eigenfunctions (both toroidal and spheroidal) and eigenfrequencies in the presence of the inferred magnetic field and recomputing the structure perturbations in this expanded set of eigenfunctions. We defer this substantial exercise to a future investigation.

In concluding, the inverse problem too comes with its own set of challenges. The signatures of magnetic-field perturbations in data are significantly weaker (with large error bars) than dominant perturbations such as flows (differential rotation) and other non-magnetic forcings \citep{schou_data}. Apart from this, \cite{Zweibel1995} showed that shifts in eigenfrequencies due to Lorentz stress can be mimicked by acoustic perturbations. Magnetic and acoustic perturbations producing the same frequency shifts can have completely different spatial structure. This might, therefore, make it difficult to attribute a frequency shift to Lorentz stress over acoustic perturbations and vice versa. Nevertheless, they also note that the distortion in eigenfunctions caused by a magnetic and an acoustic perturbation are different (unlike their signatures in eigenfrequencies).


The authors thank the anonymous referee for reviewing the study extremely carefully, suggesting numerous valuable suggestions and helping to improve the quality of the manuscript significantly. 

\appendix

\section{Linearization of MHD equations} \label{sec: MHD_linearization}

We linearize Eqns.~(\ref{eqn: MHD1})--(\ref{eqn: MHD4}) considering a static background and time-dependent first-order Eulerian perturbations in the MHD parameters $\mathbf{v}$, $p$, $\mathbf{B}$, and $\rho$:
\begin{eqnarray}
    \mathbf{v}(\mathbf{r},t) &=& \mathbf{v}_1 (\mathbf{r},t), \label{eqn: LMHD1}\\
    p(\mathbf{r},t) &=& p_0(\mathbf{r}) + p_1(\mathbf{r},t), \\
    \mathbf{B}(\mathbf{r},t) &=& \mathbf{B}_0(\mathbf{r}) + \mathbf{B}_1(\mathbf{r},t),\\
    \rho(\mathbf{r},t) &=& \rho_0(\mathbf{r}) + \rho_1(\mathbf{r},t). \label{eqn: LMHD4}
\end{eqnarray}
The background parameters $\rho_0(\mathbf{r})$ and $p_0(\mathbf{r})$ are considered to be spherically symmetric as per our choice of background model. We allow for an inhomogeneous background magnetic field $\mathbf{B}_0(\mathbf{r})$ which is not a part of model S \citep{jcd}, but shall be introduced as a subsequent perturbation. It should be noted that we adopt the Cowling approximation and therefore do not consider perturbations to the gravitational potential. Next we define a Lagrangian displacement vector field $\xiv$ that measures the displacement of a particle with respect to its background state upon introducing perturbations (with zero flow in the background $\mathbf{v}_0 = \mathbf{0}$). To first-order in $\xiv$, the velocity, or equivalently the velocity perturbation, can be written as 
\begin{eqnarray}
    \mathbf{v} \sim \mathbf{v}_1 = \partial_t \xiv.
    \label{eqn: LMHD5}
\end{eqnarray}
Substituting (\ref{eqn: LMHD1})--(\ref{eqn: LMHD5}) into Eqn.~(\ref{eqn: MHD1}) the background zeroth-order equation can be written as
\begin{equation}
    \mathbf{j}_0 \times \mathbf{B}_0 = \bnabla p_0 + \rho_0 \bnabla \phi,
\end{equation}
and to first-order in the perturbed parameters we have
\begin{eqnarray}
    \partial_t p_1 &=& -\mathbf{v}_1 \cdot \bnabla p_0 - \gamma\, p_0 \,\bnabla \cdot \mathbf{v}_1, \\
    \partial_t \mathbf{B}_1 &=& \bnabla \times (\mathbf{v}_1 \times \mathbf{B}_0), \\
    \partial_t \rho_1 &=& - \bnabla \cdot (\rho_0\, \mathbf{v}_1).
\end{eqnarray}
Replacing all instances of $\mathbf{v}_1$ in the first-order linearized equations with $\partial_t \xiv$, integrating over time, and substituting the expressions for $p_1$, $\mathbf{B}_1$, and $\rho_1$ into Eqn.~(\ref{eqn: LMHD1}) we obtain the linearized equation of motion
\begin{equation}
    \rho_0\, \omega^2 \xiv = \bnabla p_1 + \mathbf{g}\, \bnabla \cdot (\rho_0\, \xiv) + \mathbf{B}_0\times (\bnabla \times \mathbf{B}_1) - (\bnabla \times \mathbf{B}_0) \times \mathbf{B}_1,
\end{equation}
where $p_1 = -\gamma \,p_0\, \bnabla \cdot \xiv - \xiv \cdot \bnabla p_0$, and, as per the equilibrium condition, $\bnabla p_0 = \mathbf{j}_0 \times \mathbf{B}_0 - \rho_0 \bnabla \phi$. The above equation can be written as 
\begin{eqnarray}
    \rho_0\, \omega^2 \xiv &=& \cL_0 \xiv + \delta \cL \xiv, \\
    \cL_0 \xiv &=& - \bnabla (\gamma\, p_0 \,\bnabla \cdot \xiv + \rho_0\, \mathbf{g} \cdot \xiv) + \mathbf{g}\, \bnabla \cdot(\rho_0 \,\xiv), \label{eqn: linearized_MHD}\\
    \delta \cL \xiv &=& \mathbf{B}_0 \times (\bnabla \times \mathbf{B}_1) - (\bnabla \times \mathbf{B}_0) \times \mathbf{B}_1 - \bnabla [\xiv \cdot (\mathbf{j}_0 \times \mathbf{B}_0)]. \label{eqn: pert_linearized_MHD}
\end{eqnarray}
with $\mathbf{B}_1 = \bnabla \times (\xiv \times \mathbf{B}_0)$ and $\mathbf{g} = -g \,\hat{\mathbf{e}}_r$. In subsequent calculations we shall drop the subscript '0'. Any unsubscripted parameter shall be assumed to be denoting the zeroth-order background state. The above decomposition of the operator into $\cL_0$ and $\delta \cL$ allows us to define a first order perturbed state devoid of magnetic field, the eigenvalue problem for which can be defined as $\rho_0 \,\omega_0^2\, \xiv_0 = \cL_0 \xiv_0$ where $\omega_0$ and $\xiv_0$ are the magnetically unperturbed eigenfrequencies and eigenfunctions. As shown in Appendix~\ref{sec: Pert_deriv}, the operator $\delta \cL$ containing the effects of introducing a magnetic field can be introduced as a perturbation to this magnetically unperturbed operator $\cL_0$ and hence calculating the perturbations in the eigenfrequencies $\delta \omega$ and $\delta \xiv$ produced by a non-zero $\mathbf{B}$.

\section{Quasi-degenerate perturbation theory}\label{sec: Pert_deriv}

For the ease of reference, we outline the method of quasi-degenerate perturbation (and thereafter its specific case of degenerate perturbation) analysis as illustrated in the early work by \cite{lavely92}. The unperturbed wave operator is labeled $\mathcal{L}_0$ and ${}_n\omega{}_{\ell}$ is the degenerate eigenfrequency of the $2\ell + 1$ degenerate unperturbed $(n,\ell)$ multiplet. Thus, the SNRNMAIS background model satisfies the equation of motion
\begin{equation}\label{eqn: qd_eqn_motion}
     \rho_0\, {}_n\omega{}_{\ell}^2\, \boldsymbol{\xi}_{n\ell} = \mathcal{L}_0 \boldsymbol{\xi}_{n\ell}.
\end{equation}
The unperturbed eigenfrequencies ${}_n\omega{}_{\ell}$ and eigenfunctions $\boldsymbol{\xi}_{n\ell}$ are obtained by solving the above eigenvalue problem. 

For quasi-degenerate perturbation analysis, we choose a reference frequency $\omega_{\rm{ref}}$. The eigenspace $\mathcal{K}$ contains all the unperturbed multiplets $(n,\ell)$ such that $|\omega_{\rm{ref}} - {}_n\omega{}_{\ell}| < \epsilon f^2$, where $\epsilon$ is a suitably small number and $f$ governs the window around $\omega_{\rm{ref}}$ within which the eigenfrequency ${}_n\omega{}_{\ell}$ must lie for $\boldsymbol{\xi}_{n\ell}$ to be considered within the eigenspace $\mathcal{K}$. All other eigenfunctions are assumed to lie in the orthogonal subspace $\mathcal{K}^{\perp}$. Henceforth, we refer to any of the eigenmodes $(n,\ell) \in \mathcal{K}$ as $k$. When we introduce a perturbation $\mathcal{L}_0 \to \mathcal{L}_0 + \delta \mathcal{L}$ in Eqn.~(\ref{eqn: qd_eqn_motion}), the following changes are also introduced to the system:
\begin{eqnarray} 
    \omega_{k}^2 &\to& \omega_{\rm{ref}}^2 + \delta \omega^2, \label{eqn: qd1}\\
    \boldsymbol{\xi}_{k} &\to& \sum_{k \in \mathcal{K}} c_{k} \,\boldsymbol{\xi}_{k} + \delta \boldsymbol{\xi} = \boldsymbol{\xi} + \delta \boldsymbol{\xi} \label{eqn: qd3}.
\end{eqnarray}
We note that $\boldsymbol{\xi} = \sum_{k \in \mathcal{K}} c_k\, \boldsymbol{\xi}_k$ and $\delta \boldsymbol{\xi} = \sum_{k \in \mathcal{K^{\perp}}} d_k\, \boldsymbol{\xi}_k$. This implies that the corrections to the eigenfunctions arise only from the multiplets $k \in \mathcal{K}$. Forcing due to Lorentz-stress $\delta \mathcal{L \boldsymbol{\xi}}$ is given by Eqn.~(\ref{eqn: pert_linearized_MHD}). Substituting Eqns.~(\ref{eqn: qd1})--(\ref{eqn: qd3}) in (\ref{eqn: qd_eqn_motion}) and retaining first-order terms yields
\begin{eqnarray}
    &&\rho_0 (\omega_{\rm{ref}}^2 + \delta \omega^2)(\boldsymbol{\xi} + \delta \boldsymbol{\xi}) = (\mathcal{L}_0 + \delta\mathcal{L})(\boldsymbol{\xi} + \delta \boldsymbol{\xi}) \\
    &\rightarrow& (\mathcal{L}_0 - \rho_0\, \omega_{\rm{ref}}^2)\boldsymbol{\xi} - \rho_0\, \delta\omega^2\,\boldsymbol{\xi} + (-\rho_0\, \omega_{\rm{ref}}^2 + \mathcal{L}_0)\delta\boldsymbol{\xi} + \delta \mathcal{L}\boldsymbol{\xi} = 0. \label{eqn: qd_first_ord}
\end{eqnarray}
Expanding out $\boldsymbol{\xi}$ in the eigenspace spanned by $\mathcal{K}$ as: $\mathcal{L}_0 \boldsymbol{\xi} = \mathcal{L}_0 \sum_k c_k\, \boldsymbol{\xi}_k = \sum_k c_k \,\rho_0 \,\omega_k^2\, \boldsymbol{\xi}_k$ and thereafter taking an inner product with $\boldsymbol{\xi}_{k'}$, where $k' \in \mathcal{K}$, and integrating over the solar volume we obtain
\begin{eqnarray}
    \braket{\boldsymbol{\xi}_{k'}|\mathcal{L}_0\boldsymbol{\xi}} = \sum_k c_k \,\omega_k^2 \braket{\boldsymbol{\xi}_{k'}|\rho_0\,\boldsymbol{\xi}_k} = \sum_k c_k\, \omega_k^2\, \delta_{k'k}.
\end{eqnarray}
On applying similar inner products to the rest of Eqn.~(\ref{eqn: qd_first_ord}) we find
\begin{equation}
    \braket{\boldsymbol{\xi}_{k'}|(\mathcal{L}_0 - \rho_0\, \omega_{\rm{ref}}^2)\boldsymbol{\xi}} - \delta\omega^2\braket{\boldsymbol{\xi}_{k'}|\rho_0\,\boldsymbol{\xi}} + \braket{\boldsymbol{\xi}_{k'}|( \mathcal{L}_0-\rho_0\, \omega_{\rm{ref}}^2 )\delta \boldsymbol{\xi}} + \braket{\boldsymbol{\xi}_{k'}|\delta \mathcal{L}\boldsymbol{\xi}} = 0.
\end{equation}

Recalling that $\boldsymbol{\xi}_{k'}$ and $\delta \boldsymbol{\xi}$ exist in orthogonal eigenspaces the equation simplifies to
\begin{equation}
    \sum_k c_k \,\omega_k^2\, \delta_{k'k} - \omega^2_{\rm{ref}} \sum_k c_k\, \delta_{k'k} - \sum_k \delta\omega^2 \,c_k\,\delta_{k'k} + 0 + 0 + \sum_k\, c_k \Lambda_{k'k} = 0.
\end{equation}
This may be cast into the form of an eigenvalue problem, namely,
\begin{equation}
    \sum_k [\Lambda_{k'k} - (\omega_{\rm{ref}}^2 - \omega_k^2)\, \delta_{k'k} ]\,c_k =  \delta\omega^2\, c_{k'},
\end{equation}
or, equivalently,
\begin{equation} \label{eqn: general_Z_eigenproblem}
    \sum_{k \in \mathcal{K}} \mathcal{Z}_{k'k}\,c_k  = \delta \omega^2 \,c_{k'}.    
\end{equation}
In the nomenclature of \cite{lavely92} $\mathcal{Z}_{k'k} = \Lambda_{k'k} - (\omega_{\rm{ref}}^2 - \omega_k^2)\, \delta_{k'k}$ are elements of their supermatrix and $\Lambda_{k'k} = \braket{\boldsymbol{\xi}_{k'}|\delta \mathcal{L}\boldsymbol{\xi}_k}$ is their general matrix element or the coupling matrix.

Finally, as a corollary of the theory developed above, let us consider the case of $\cK$ consisting of modes that are all degenerate with frequency $\omega_0$. In this case, setting $\omref = \omega_0$, we see that $Z_{k'k}$ reduces to $\Lambda_{k'k}$. Now the leading order corrections to eigenfrequency and eigenfunction are described by the eigenvalues and eigenvectors of $\Lambda_{k'k}$: 
\begin{equation} \label{eqn:DPT}
    \sum_{k \in \mathcal{K}} \Lambda_{k'k}\,c_k  = \delta \omega^2 \,c_{k'}    .
\end{equation}
This is now degenerate perturbation analysis. \cite{ritzwoller} lays out a detailed formulation of this method for the case of a perturbation in the form of differential rotation on the background SNRNMAIS model.

\section{The $\delta \mathcal{L}$ operator and coupling matrix for Lorentz stress} \label{sec: deriv_cpmat}

In the regime of linearized ideal MHD with small perturbations about an equilibrium \citep{goedbloed2004}, the perturbation operator due to magnetic fields in a non-magnetic background is given by (\ref{eqn: pert_linearized_MHD})
\begin{equation}
    \delta \cL \xiv = \frac{1}{4\pi} \Big( \mathbf{B} \times \{\grad \times [\bnabla \times (\xiv \times \mathbf{B})]\} - (\grad \times \mathbf{B}) \times [\bnabla \times (\xiv \times \mathbf{B})] - \bnabla \{\xiv \cdot [(\bnabla \times \mathbf{B}) \times \mathbf{B}]\}\Big).
\end{equation}

We compute the coupling matrix $\Lambda_{k'k} = \braket{\boldsymbol{\xi}_{k'}|\delta \mathcal{L} \boldsymbol{\xi}_{k}}$ exclusively for terms involving magnetic fields. As elucidated in Section 6.2.3 of \cite{goedbloed2004}, one ends up with the following terms:
\begin{eqnarray} \label{eq: G&P_expr}
    \Lambda_{k'k} &=&\frac{1}{4\pi} \int_{V_{\odot}} \mathrm{d}^3\mathbf{r} \left[ \mathbf{Q} \cdot \mathbf{R} + \tfrac{1}{2}\, \bnabla p_B \cdot (\xiv_{k} \bnabla \cdot \xiv_{k'}^* + \xiv_{k'}^* \bnabla \cdot \xiv_k) + \tfrac{1}{2}\,\mathbf{j} \cdot (\xiv_k \times \mathbf{R} + \xiv_{k'}^* \times \mathbf{Q}) \right]  \\ 
    &-& \frac{1}{4\pi}\int_{\sum_{\odot}} d\Sigma \,\hat{\mathbf{n}} \cdot \left[ \xiv_{k'}^* \,(\xiv_k \cdot \bnabla p_B - \mathbf{B} \cdot \mathbf{Q}) + \mathbf{B}\, \xiv_{k'}^* \cdot \mathbf{Q} + \tfrac{1}{2}\,\mathbf{j} \,\mathbf{B} \cdot (\xiv_k \times \xiv_{k'}^*) - \tfrac{1}{2}\,(\mathbf{j} \times 
    \mathbf{B}) \cdot (\xiv_k\, \xiv_{k'}^* - \xiv_{k'}^*\,\xiv_k)\right], \nonumber
\end{eqnarray}
where $\mathbf{Q} = \bnabla \times (\xiv_k \times \mathbf{B}), \mathbf{R} = \bnabla \times (\xiv_{k'}^* \times \mathbf{B})$ and $\bnabla p_B = \mathbf{j} \times \mathbf{B}$. We carry out a tedious exercise of isolating the $\Bv\Bv$ terms from the volume integral in Eqn.~(\ref{eqn: cp_mat_corr}) until we end up with the following expression for the volume integral term in coupling matrix:
\begin{eqnarray} \label{eqn: cp_mat_corr}
    \Lambda_{k'k} &=& \frac{1}{4\pi} \int_{V_{\odot}} \mathrm{d}^3r\, \cH : \left\{\tfrac{1}{2}\left[\bnabla \xiv_k \cdot (\grad \xiv_{k'}^*)^T + \bnabla \xiv_{k'}^* \cdot (\bnabla \xiv_k)^T \right] + \tfrac{1}{2}\left( \bnabla \xiv_{k'}^* \cdot \bnabla \xiv_k + \bnabla \xiv_k \cdot \bnabla \xiv_{k'}^*\right) + \boldsymbol{I}\, \bnabla \cdot \xiv_{k'}^* \bnabla \cdot \xiv_k \right. \nonumber\\
    &+& \left. \tfrac{1}{2}\left( \xiv_{k'}^*\cdot \bnabla \bnabla \xiv_k + \xiv_k \cdot \bnabla \bnabla \xiv_{k'}^* \right) 
    - \tfrac{1}{2}\left( \xiv_k \bnabla \bnabla \cdot \xiv_{k'}^* + \xiv_{k'}^* \bnabla \bnabla \cdot \xiv_k \right) - \tfrac{3}{2} \left( \bnabla \xiv_k \bnabla \cdot \xiv_{k'}^* + \bnabla \xiv_{k'}^* \bnabla \cdot \xiv_k \right)  \right\} 
\end{eqnarray}
In obtaining the above expression for $\Lambda_{k'k}$ from Eqn.~(\ref{eq: G&P_expr}), we accumulated the following boundary terms
\begin{equation} \label{eq: mag_surface_terms}
   \frac{1}{4\pi} \int_{\Sigma_{\odot}} d \Sigma \,\hat{\mathbf{n}} \cdot \big[ \xiv^*_{k'}\cdot(\bnabla \, \Bv) \, \Bv \cdot \xiv_{k} + \Bv \Bv \cdot \xiv_{k} (\bnabla \cdot \xiv^*_{k'}) - \tfrac{1}{2}\Bv \Bv : \bnabla(\xiv_{k} \, \xiv^*_{k'})\big]+ (\xiv_k \leftrightarrow \xiv^*_{k'})
\end{equation}
For model S \citep{jcd}, the top boundary is at $r_S = R_{\odot} + 0.5$Mm. For trapped normal modes of the Sun, the upper turning points all lie below the photosphere; above this, waves become evanescent and their eigenfunctions decay into the atmosphere. The dominant sensitivities of these modes is therefore firmly focused in the solar interior. Additionally, modes are primarily sensitive to changes in the speed of propagation - which in this case is the Alfven speed \citep{hanasoge12_mag}. As a consequence, the proper way to parametrise the inversion is to use $\cH/\rho$, since this represents the Alfven-speed squared and hence corresponding effective kernels are $\rho \cB$. Taking these two factors into account, we neglect the boundary terms - which are evaluated at a layer above the upper turning points and whose contribution is therefore small in comparison to the bulk integrals, which possess the predominant sensitivity. Nevertheless, we would like to caution the reader about the fact that fields in the solar atmosphere affect mode frequencies - waves propagating in magnetized regions undergo mode conversion \citep{Cally_Bogdan_93,cally97} and these magneto-acoustic waves escape into the atmosphere via field lines. This can create shifts in mode frequencies \citep{wright92} and lead to increased linewidths as well \citep{chaplin00,komm00,pinter08}. To properly account for these effects, the eigenfunction basis needs to admit magneto-acoustic modes - such a basis will contain both spheroidal and toroidal eigenfunctions. In the present treatment, the background is hydrodynamic and therefore only allows spheroidal (acoustic) eigenfunctions. As a consequence, the kernels obey classical acoustic-mode physics and are therefore dominantly sensitive to the interior alone. In more sophisticated models, the background would contain a magnetic field, the eigenfunctions would comprise both spheroidal and toroidal modes - the kernels would then show significant sensitivity to both the interior (near-surface layers dominating) and the field in the atmosphere.

In Eqn.~(\ref{eqn: cp_mat_corr}), clearly $\Lambda_{k'k}^{\dagger} = \Lambda_{k'k}$ and therefore the coupling-matrix (or equivalently the supermatrix $\mathcal{Z}_{k'k}$) is Hermitian. This ensures that the corrections to the eigenfrequencies are real and hence the perturbed solutions are temporally stable. Expanding out each of the vectors and tensors in Eqn.~(\ref{eqn: cp_mat_corr}) in the GSH basis and evaluating the surface integral, we are left with an expression that may be cast in the form shown in Eqn.~(\ref{eqn: lambda_m}), where $B_{st}^{\mu \nu}$ are sensitivity kernels corresponding to Lorentz-stress components $h_{st}^{\mu \nu}$.

\subsection{Sensitivity Kernels}\label{sec:mag_kern}
The coupling-matrix element is given as an integral transform over $\cH$ by
\begin{equation}\label{lor_kernel_intro}
\LamB_{k'k} = \inner{\xiv_{k'}}{\dLB\xiv_{k}}=\solint r^2 \sum_{\substack{st \\ \mu\nu}} \cB_{st}^{\mu\nu}(r)\, h_{st}^{\mu\nu}(r),
\end{equation}
where $\cB_{st}^{\mu\nu}$ are eigenfunction-dependent magnetic-sensitivity kernels. The prescription for evaluating these kernels and explicit expressions may be found in \citet{hanasoge17}. The coupling integral $\inner{\xiv_{k'}}{\dLB\xiv_{k}}$, which has been reduced to the radial integral form obtained in Eqn.~(\ref{lor_kernel_intro}) contains no boundary terms. It is indeed the case that the magnetic field is assumed to vanish at the surface in this analysis. Relaxing this assumption will introduce boundary terms, which involve integrals only over the solar surface. Since $h_{st}^{\mu\nu}$ is symmetric in the interchange of $\mu$ and $\nu$, we ascribe the same symmetry to $\cB_{st}^{\mu\nu}$ as well, without loss of generality.

Using the Mathematica package developed for this work \citep{GSH_repo}, we automate the manipulation of tensor spherical harmonics via the method of GSHs and obtain explicit forms of the Lorentz-stress sensitivity kernels $\cB_{st}^{\mu\nu}$. Each tensorial component of these kernels can be written in the form
\begin{equation}
    \cB_{st}^{\mu\nu}  = 4\pi (-1)^{m'}\gamma_{\ell'} \gamma_{s} \gamma_{\ell}\wigred{-m'}{t}{m}\cG_{s}^{\mu\nu}, \label{eqn: m_ind_kern}
\end{equation}
where the four independent components of $\cG_{st}^{\mu\nu}$ have the following closed-form expressions:
\begin{eqnarray}
\cG_{s}^{--} = &&\frac{-1}{2r^2}  \Bigg[\wigred{2}{-2}{0} \chi_1^{--}(k,k') +\wigred{0}{-2}{2} \chi_1^{--}(k',k) \nonumber\\ &&\hspace{1em} +~\wigred{1}{-2}{1} \big\{ \chi_2^{--}(k,k') + \chi_2^{--}(k',k) \big\} +~\wigred{3}{-2}{-1} \chi_3^{--}(k,k') + \wigred{-1}{-2}{3} \chi_3^{--}(k',k)\Bigg],\label{eq:kern_mm} \\
\cG_{s}^{0-} = &&\frac{1}{4r^2} \Bigg[\wigred{1}{-1}{0} \chi_1^{0-}(k,k') + \wigred{0}{-1}{1}\chi_1^{0-}(k',k) \nonumber\\ &&\hspace{6em} +~\wigred{-1}{-1}{2} \chi_2 ^{0-}(k,k') +~\wigred{2}{-1}{-1}\chi_2^{0-}(k',k)\Bigg],\label{eq:kern_0m} \\
\cG_{s}^{00} = &&\frac{1}{2r^2} (1+p) \bigg\{\tfrac{1}{2}\,\wigred{0}{0}{0} [\chi_1^{00}(k,k') + \chi_1^{00}(k',k) ] + \wigred{-1}{0}{1} [\chi_2^{00}(k,k') + \chi_2^{00}(k',k) ] \bigg\}, \nonumber\\\label{eq:kern_00} \\
\cG_{s}^{+-} = &&\frac{1}{4r^2} (1+p) \bigg\{\tfrac{1}{2}\,\wigred{0}{0}{0} [\chi_1^{+-}(k,k') + \chi_1^{+-}(k',k)] \nonumber \\ &&\hspace{6em} +~\wigred{-2}{0}{2} [\chi_2^{+-}(k,k') + \chi_2^{+-}(k',k) ]  +~\wigred{-1}{0}{1}  [\chi_3^{+-}(k,k') +  \chi_3^{+-}(k',k) ] \bigg\}. \label{eq:kern_pm}
\end{eqnarray}
where $p \equiv (-1)^{\ell'+\ell+s}$ and
\begin{eqnarray}
\chi_1^{--}(k,k') &=& \om{\ell'}{0}\,\om{\ell'}{2}\big[V'(3U - 2\om{\ell}{0}^2V + 3r\dot{U}) - rU\dot{V}' \big],\\
\chi_2^{--}(k,k') &=& \om{\ell'}{0}\,\om{\ell}{0}\big(3UV' - 2\om{\ell'}{0}^2V'V + \om{\ell'}{2}^2V'V + rV\dot{U}' - rU\dot{V}' - U'U \big),\\
\chi_3^{--}(k,k') &=& \om{\ell}{0}\,\om{\ell'}{0}\,\om{\ell'}{2}\,\om{\ell'}{3}V'V,\\
\chi_1^{0-}(k,k') &=& \om{\ell'}{0}\big\{4 \om{\ell}{0}^2 V'V + U'(8U - 5\om{\ell}{0}^2 V) - 3r \om{\ell}{0}^2 V\dot{V}' + 2r\dot{U}\dot{V}' - r\om{\ell}{0}^2 V'\dot{V}  + r^2 V' \ddot{U} \nonumber \\ &+& U\,[(\om{\ell}{0}^2 -2\om{\ell'}{0}^2 -6)V'  + r(4\dot{V}' - r\ddot{V}' ] \big\}, \\
\chi_2^{0-}(k,k') &=& \om{\ell}{0}\, \om{\ell'}{0}\, \om{\ell}{2} \big[UV' + V(U' - 4V' + 3r\dot{V}') + rV'\dot{V} \big], \\
\chi_1^{00}(k,k') &=& 2\big(-2rU\dot{U}' + \om{\ell}{0}^2 rV\dot{U}' - 5 \om{\ell'}{0}^2 V'U + 2 \om{\ell}{0}^2 \om{\ell'}{0}^2 V'V + \om{\ell'}{0}^2 rU\dot{V}' + 3U'U \ \big), \\
\chi_2^{00}(k,k') &=& \mbox{}-\om{\ell'}{0}\,\om{\ell}{0}\big(-U'V + V'V + rV\dot{U}' - 2rV\dot{V}' + rU\dot{V}' + r^2\dot{V}'\dot{V} \big), \\
\chi_1^{+-}(k,k') &=& 2\big(-2r\dot{U}'U + \om{\ell}{0}^2 r\dot{U}'V - r^2 \dot{U}'\dot{U} - \om{\ell'}{0}^2UV' + \om{\ell'}{0}^2rU\dot{V}' + U'U \big), \\
\chi_2^{+-}(k,k') &=& \mbox{}-2\om{\ell}{0} \,\om{\ell'}{0}\, \om{\ell}{2} \om{\ell'}{2} V'V, \\
\chi_3^{+-}(k,k') &=& \om{\ell}{0}\om{\ell'}{0}\big(-rV\dot{U}' -V'U + rU\dot{V}' + U'U \big).
\end{eqnarray}
Here $U,V \equiv {}_nU{}_{\ell},{}_nV{}_{\ell}$, and $U',V' \equiv {}_{n'}U{}_{\ell'},{}_{n'}V_{\ell'}$ and $\dot{U} = \partial_r\, {}_nU{}_{\ell}, \dot{V} = \partial_r\, {}_nV{}_{\ell}, \dot{U'} = \partial_r\, {}_{n'}U{}_{\ell'}$ and $\dot{V'} = \partial_r\, {}_{n'}V{}_{\ell'}$.

The following selection rules are imposed on the kernels:
\begin{itemize}
    \item $|l'-l| \leq s \leq |l'+l|$.
    \item $m-m'+t=0$.
    \item $\mathcal{B}_{st}^{--}=0$ for s = 0,1.
    \item $\mathcal{B}_{st}^{0-}=0$ for s = 0. 
\end{itemize}
Kernel components $\cB_{st}^{\mu\nu}$ are found to have these following properties:
\begin{enumerate}
\item $\cB_{st}^{\mu\nu} = \cB_{st}^{\nu\mu}$ (by construction)
\item $\cB_{st}^{--} = (-1)^{\ell+\ell'+s}\,\cB_{st}^{++}$
\item $\cB_{st}^{0-} = (-1)^{\ell+\ell'+s}\,\cB_{st}^{+0}$
\item $\cB_{st}^{00} = \cB_{st}^{+-}=\cB_{st}^{-+}=0$ for odd $\enc{\ell'+\ell+s}$
\item $\cB_{st}^{\mu\nu\,*}(k,k') = (-1)^{\ell'+\ell+s+t}\,\cB_{s\bar{t}}^{\mu\nu}(k',k)$
\end{enumerate}

Self-coupling sensitivity kernels for axisymmetric magnetic fields exhibit the useful property that $B_{s0}^{\mu\nu}(m)$ may be neatly separated into $m$-dependent and $m$-independent terms
\begin{eqnarray}
    B_{s0}^{\mu\nu}(m) &=& (-1)^m\wigfull{\ell}{s}{\ell}{-m}{0}{m} \tilde{\mathcal{K}}_{s}^{\mu\nu} \label{eqn:kern_m_dep} \\
    \text{using the relation} \quad \wigfull{\ell}{s}{\ell}{-m}{0}{m} &=& \frac{(-1)^{m+\ell}}{\ell} \>\cP_s(m)\, \wigfull{\ell}{s}{\ell}{-\ell}{0}{\ell} \label{eqn:wigner_P_ljm_relation}\\
    B_{s0}^{\mu\nu}(m) &=& \mathcal{P}_{s}^{(l)}(m)\, \mathcal{K}_{s}^{\mu\nu}, \label{eqn:Kernel_m_separation}
\end{eqnarray}
where $\mathcal{K}_{s}^{\mu\nu} = \frac{(-1)^{\ell}}{\ell} \wigfull{\ell}{s}{\ell}{-\ell}{0}{\ell} \tilde{\mathcal{K}}_{s}^{\mu\nu} $. As shown in Eqn.~(\ref{eqn:a_coeff_kern}), the relation~(\ref{eqn:Kernel_m_separation}) is essential in cutting down on the calculation of $(2\ell+1)$ kernels for each $(n,\ell)$ multiplet. Thus, it suffices to compute a single kernel $\mathcal{K}_{s}^{\mu\nu}$ for every multiplet.
\begin{figure}
\centering
    \includegraphics[width=\textwidth]{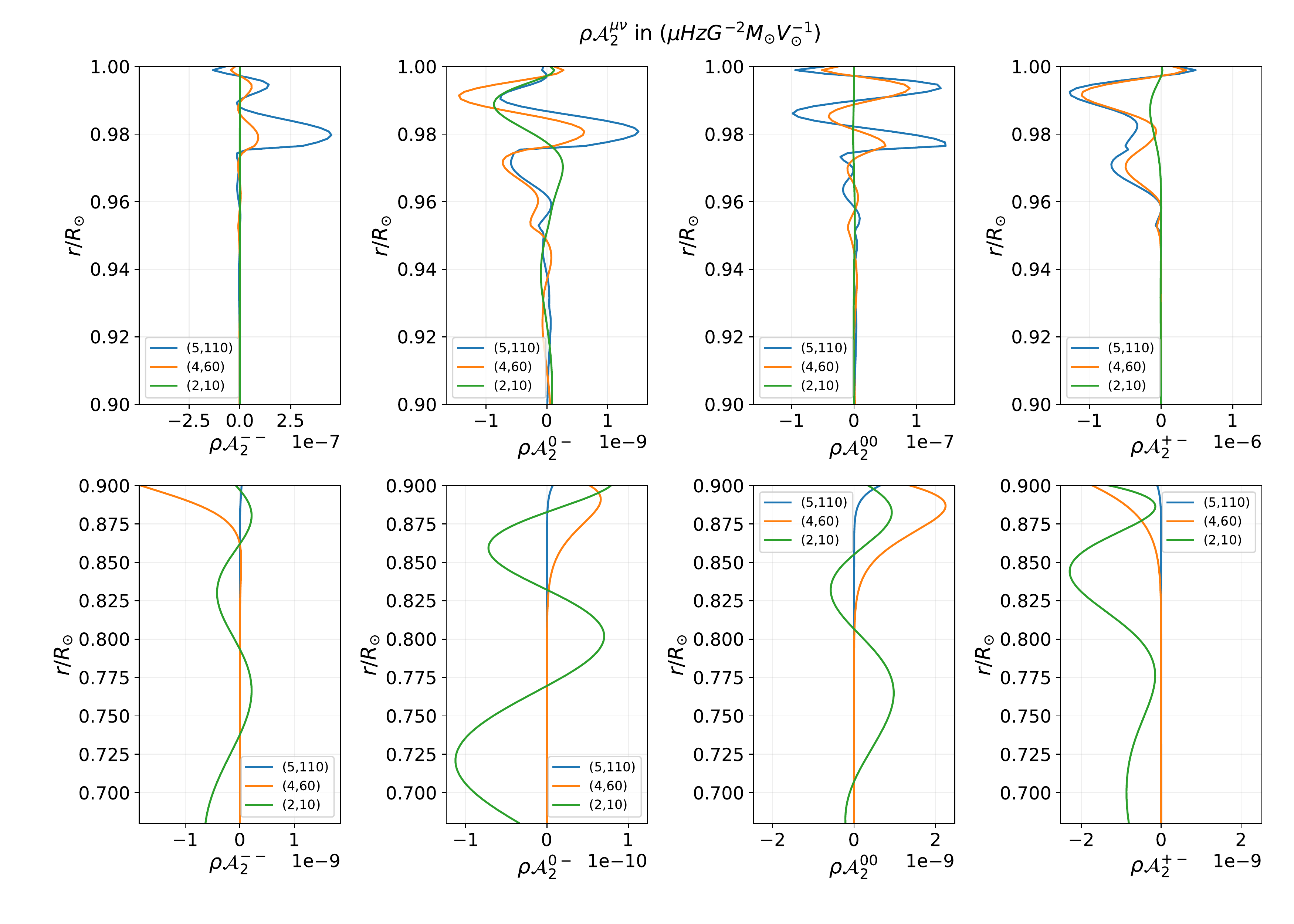}
    \caption{Density-scaled Lorentz stress kernels for modes ${}_{2}\rm{S}_{10}, {}_{4}\rm{S}_{60}$ and ${}_{5}\rm{S}_{110}$. The kernels in the \textit{top} row extends between $0.9R_{\odot}<r<R_{\odot}$ and shows the strong near-surface sensitivity of ${}_{5}\rm{S}_{110}$ and ${}_{4}\rm{S}_{60}$ but weaker sensitivity of ${}_{2}\rm{S}_{10}$. The kernels in the \textit{bottom} row are plotted from $0.9 R_{\odot}$ down to the tachocline and show the dominance of sensitivity of the ${}_{2}\rm{S}_{10}$ mode over ${}_{5}\rm{S}_{110}$ and ${}_{4}\rm{S}_{60}$ in deeper layers.
    }
    \label{fig:110_dom_deeper}
\end{figure}

\section{Custom magnetic fields} 
\subsection{Construction of $\Bv$}\label{sec:B_construction}
Using identities $\grad_1 Y_{\ell m} = \om{\ell}{0} \enc{Y_{\ell m}^{-1} \ev{-} + Y_{\ell m}^{+1} \ev{+}}$, $\ev{r} \times\grad_1 Y_{\ell m} = i\,\om{\ell}{0} \enc{Y_{\ell m}^{-1} \ev{-} - Y_{\ell m}^{+1} \ev{+}}$ and $Y_1^0(\theta,\phi) = \gam{1}\cos\theta$, we obtain the following expressions:
\begin{enumerate}
    \item  A toroidal field $\Bv = \alpha(r) \ev{r} \times\grad_1 Y_{\ell m} = - \alpha(r) \sin\theta \ev{\phi}$ may be given as $B_{10} = i \alpha(r) / \gam{1} \enc{-1,0,1}$, with all other $B_{st}$ components being $0$. $\alpha(r)$ represents a function in radial distance $r$ which may be tuned to construct a region containing a dominant, insignificant or intermediate-magnitude toroidal magnetic field.
    \item A dipolar field $\Bv = \beta(r) (2\cos\theta \ev{r} + \sin\theta\ev{\theta})$ with $\beta \propto r^{-3}$ is given as $B_{10} = -\beta(r)/\gam{1} \enc{1,-2,1}$, with all other $B_{st}$ components being $0$. 
\end{enumerate}
We choose to express a vector in GSH basis $\Bv = \enc{a Y_{\ell m}^{-1} \ev{-} + b Y_{\ell m}^0 \ev{0} + c Y_{\ell m}^{+1} \ev{+}}$ in the convenient form $\Bv = \enc{a,b,c}$. In the Sun, the tachocline is believed to contain a dominantly toroidal field, while the simplest magnetic configuration at the surface is a typical dipole (during solar minima \cite{Andres13,Bhowmik2018}). Therefore, a crude approximation would involve constructing a global magnetic field that changes from a toroidal to a dipolar configuration as a function of radial distance. This forms the motivation for the next mathematical exercise that takes purely toroidal and dipolar magnetic fields and constructs a mixed field,
\begin{equation} \label{eqn:init_mixed_B}
    \Bv_{10}^{\rm{mixed}} (r) = -i\> \frac{\alpha(r)}{\gam{1}} \gshvec{1}{0}{-1}  - \frac{\beta(r)}{\gam{1}} \gshvec{a}{-2b}{a}.
\end{equation}
The above form of $\Bv_{10}^{\rm{mixed}} (r)$ is chosen with the following motivation
\begin{itemize}
    \item $\Bv_{10}^{\rm{mixed}} (r)$ has to satisfy the constraint $\bnabla\cdot\Bv = 0$. In the GSH basis, this translates to
    \begin{equation} \label{eqn:divBGSH}
      B^{-}_{10} + B^{+}_{10} = r^{-1}\partial_r (r^2 B_{10}^0).
    \end{equation}
    \item We write $\Bv_{10}^{\rm{mixed}}(r) = \Bv_{10}^{\rm{tor}}(r) + \widetilde{\Bv}_{10}(r)$, where $\Bv_{10}^{\rm{tor}}(r)$ represents the toroidal part (first term in Eqn.~\ref{eqn:init_mixed_B}) and $\widetilde{\Bv}_{10}(r)$ represents the remnant field, excluding the toroidal part (second term in Eq.~[\ref{eqn:init_mixed_B}]). Invoking the linearity of the Eqn.~(\ref{eqn:divBGSH}), we write
    \begin{equation}
        (B_{10}^{-,\rm{tor}} + B_{10}^{+,\rm{tor}}) + (\widetilde{B}_{10}^{-} + \widetilde{B}_{10}^{+}) = r^{-1}\partial_r (r^2 B_{10}^{0,\rm{tor}}) + r^{-1}\partial_r (r^2 \widetilde{B}_{10}^0).  
    \end{equation}
    The toroidal term $\Bv_{10}^{\rm{tor}}(r)$ is eliminated from this equation and therefore stated separately in Eqn.~(\ref{eqn:init_mixed_B}) since this involves no further effort to satisfy the divergence-free condition on field. 
    \item This simplifies to solving for the Ordinary Differential Equation (ODE)
    \begin{equation}
        \widetilde{B}_{10}^{-} + \widetilde{B}_{10}^{+} = r^{-1}\partial_r (r^2 \widetilde{B}_{10}^0).
    \end{equation}
    Substituting the second term from Eqn.~(\ref{eqn:init_mixed_B}) in the above equation leads us to an ODE in $r$ that relates $a(r)$ and $b(r)$,
    \begin{equation}
        a(r) = b(r) - r\, \dot{b}(r). \label{eqn: a_b_reln}
    \end{equation}
\end{itemize}
We obtain the following final form of $\Bv$
\begin{equation}\label{eqn: mag_field}
B_{st}(r) = 
\begin{cases}
-i\> \frac{\alpha(r)}{\gam{1}} \gshvec{1}{0}{-1}  - \frac{\beta(r)}{\gam{1}} \gshvec{b-r\dot{b}}{-2b}{b-r\dot{b}}, & \text{for} (s,t) = (1,0) \\
0, & \text{for} (s,t) \neq (1,0),
\end{cases}
\end{equation}
where $b(r)=1$ (or equivalently, any real constant) where field is perfectly dipolar. The term $r\dot{b}(r)$ appears as a consequence of fixing the divergence to zero and is only nonzero in the transition region where $b(r)$ goes from $0$ to $1$. It may be verified -- using $\grad \cdot \Bv = g_{\alpha\beta} (\grad\Bv)^{\alpha\beta}$ \citep[][]{GSH_repo} -- that the two parts in (\ref{eqn: mag_field}) (toroidal and dipolar) satisfy the solenoidal condition independently.

\begin{figure}
\centering
    \includegraphics[totalheight=12cm]{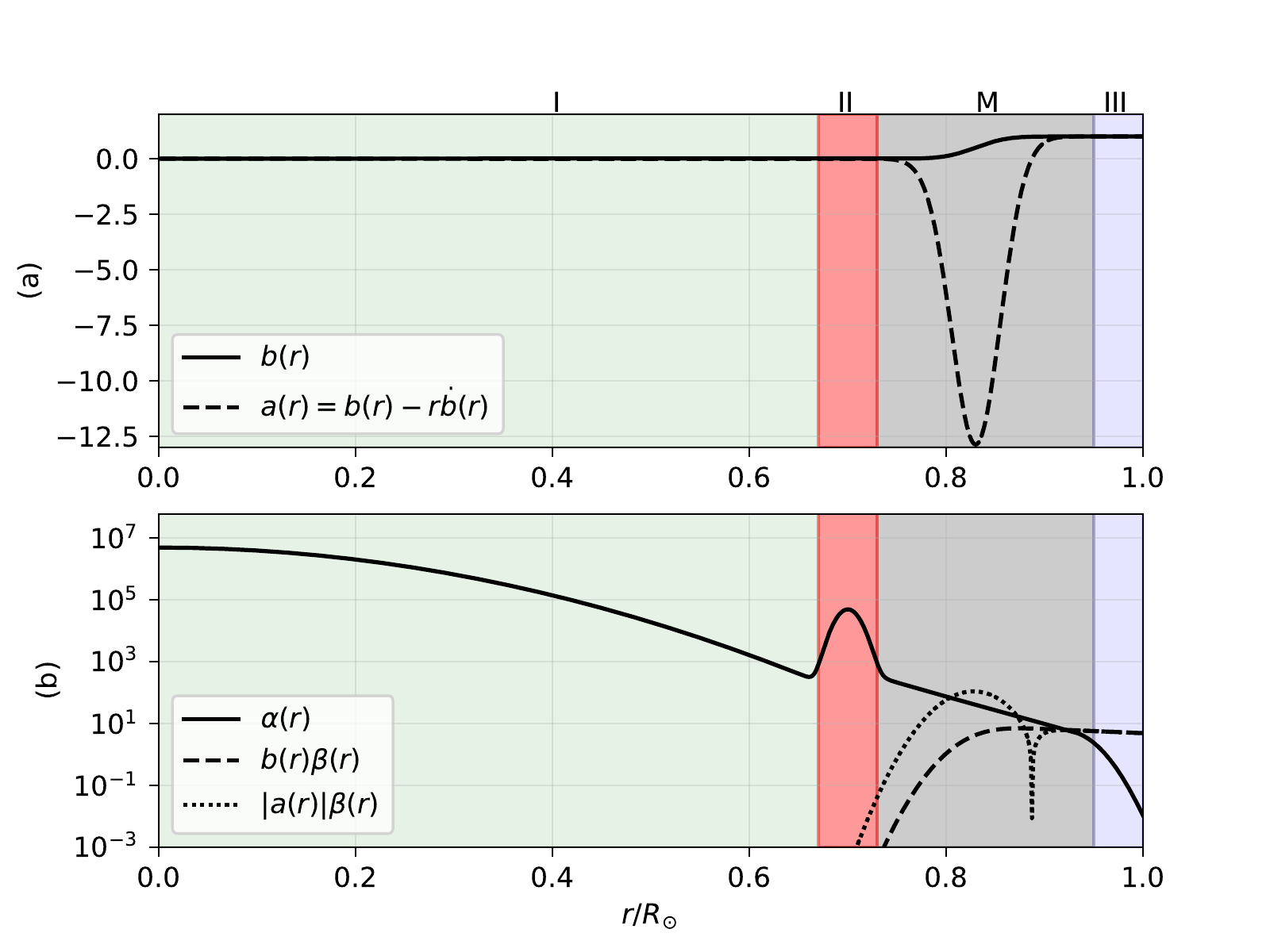}
    \caption{(a) The radial profiles of parameters $a$ and $b$ in Eqn.~(\ref{eqn: a_b_reln}). The smooth transition of $b$ from $0$ to $1$ is modelled using a sigmoid around $r=0.7 R_\odot$. (b) Strength of the three component of total $\Bv$ in Eqn.~(\ref{eqn: mag_field}). $\alpha(r)$ is the sum of two Gaussians, one centred at $r=0$ with peak $10^7 G$ and another at $r=0.7 R_{\odot}$ with peak $10^5 G$ and an extended tail.  The $r=0.7 R_\odot$ mark is roughly where the tachocline is located and hence the peak in the toroidal component $\alpha(r)$. Figure~(b) shows the poloidal (dipolar) field $\beta$ starting to dominate over the toroidal field by at least three orders of magnitude for $r\gtrsim 0.95 R_\odot$.}
    \label{fig:field_params}
\end{figure}

\subsection{Construction of $\boldsymbol{\mathcal{H}}$}\label{sec: H_expr}

The coupling matrix $\Lambda_{m'm}$ requires the expansion of the Lorentz stress $\boldsymbol{\mathcal{H}}$ in the basis of GSH,
\begin{eqnarray} 
    \boldsymbol{\mathcal{H}} &=& \sum_{s t} h_{s t}^{\mu \nu} Y^{\mu + \nu}_{s t} \hat{e}_{\mu} \hat{e}_{\nu} = \sum_{s_1,t_1}\sum_{s_2,t_2} B^{\mu}_{s_1 t_1} B^{\nu}_{s_2 t_2} Y^{\mu}_{s_1 t_1} Y^{\nu}_{s_2 t_2} \hat{e}_{\mu} \hat{e}_{\nu} \\
    h^{\mu \nu}_{s t} &=& \sum_{s_1,s_2,t_1,t_2} B_{s_1 t_1}^{\mu} B_{s_2 t_2}^{\nu} \int Y^{* \mu + \nu}_{s t} Y_{s_1 t_1}^{\mu} Y_{s_2 t_2}^{\nu} \mathrm{d}\Omega \\
    &=& \sum_{s_1,s_2,t_1,t_2}B^{\mu}_{s_1 t_1}B^{\nu}_{s_2 t_2} (-1)^{\mu + \nu + t} \sqrt{\frac{(2s+1)(2s_1 + 1)(2s_2 + 1)}{4 \pi}} \tj{s_1}{s}{s_2}{\mu}{-(\mu+\nu)}{\nu} \tj{s_1}{s}{s_2}{t_1}{-t}{t_2}, \label{eqn: generic_Hmunu}
\end{eqnarray}
where $\mu$ or $\nu$ = $\{-,0,+\}$. Given that we have $\boldsymbol{B}_T = B^{\mu}_{s_0 t_0} Y^{\mu}_{s_0 t_0} \hat{e}_{\mu}$, the expression for $h^{\mu \nu}_{s t}$ becomes
\begin{eqnarray}\label{eqn:h_B_relation}
    h_{st}^{\mu\nu} = B^{\mu}_{s_0 t_0}\,B^{\nu}_{s_0 t_0}\, (-1)^{\mu + \nu + t} (2s_0+1)\, \sqrt{\frac{2s + 1}{4 \pi}} \tj{s_0}{s}{s_0}{\mu}{-(\mu+\nu)}{\nu} \tj{s_0}{s}{s_0}{t_0}{-t}{t_0}.
\end{eqnarray}
For the axisymmetric magnetic field constructed in Section~\ref{sec:B_construction}, we have set $s_0=1$ and $t_0=0$. Wigner-$3j$ selection rules \cite[Appendix C of][]{DT98} dictate that $\cH$ only have $s=0,1,2$ and $t=0$. Then we have the form
\begin{equation}
h_{s0}^{\mu\nu} = 3 \gam{s}\, B^{\mu}_{10}\,B^{\nu}_{10}\, (-1)^{\mu + \nu}\, \tj{1}{s}{1}{\mu}{-(\mu+\nu)}{\nu} \tj{1}{s}{1}{0}{0}{0}.
\end{equation}
But we know that $\wigfull{1}{s}{1}{0}{0}{0}$ vanishes for odd s. Thus $\cH$ has no $s=1$ and has non-zero $s=0$ components, which is different from how differential rotation couples modes. The $s=0$ feature of the Lorentz stress tensor indicates a net shift from the unperturbed mode frequency ${}_n\omega{}_{\ell}$ for a particular multiplet $\mode{n}{l}$ as this term couples with $\wigfull{\ell '}{0}{\ell}{-m}{0}{m}$, which is independent of $m$.

In Section~\ref{Sec: Results}, we carry out calculations for the forward problem to estimate the frequency splitting induced by a synthetic magnetic field that remains toroidal from the core to the tachocline, subsequently transitioning to mixed and finally to purely dipolar near the solar surface.

\section{Transforming $\boldsymbol{\mathcal{H}}$ from GSH space to physical space} \label{sec: GSH2SPHR}
The Lorentz-stress sensitivity kernels have four independent components (see symmetry relations of $\mathcal{B}_{st}^{\mu\nu}$ in Appendix~\ref{sec:mag_kern}). The six independent Lorentz stress components in $(r,\theta,\phi)$ space may be related to the ($0,+,-$) space of GSH through the following relations
\begin{eqnarray}
    B_r B_r &=& \sum_{s,t} h_{st}^{00} Y_{st}^0, \label{eqn: BrBr_App}\\
    B_r B_{\theta} &=& \tfrac{1}{\sqrt{2}}\,\sum_{s,t} (h_{st}^{0-} Y_{st}^{-} - h_{st}^{0+}Y_{st}^{+}), \\
    B_r B_{\phi} &=& -\tfrac{i}{\sqrt{2}}\, \sum_{s,t} (h_{st}^{0-}Y_{st}^{-} + h_{st}^{0+}Y_{st}^{+}), \\
    B_{\theta} B_{\theta} &=& \tfrac{1}{2}\,\sum_{s,t}(h_{st}^{++}Y_{st}^{+2} - 2 h_{st}^{+-}Y_{st}^{0} + h_{st}^{--}Y_{st}^{-2}), \label{eqn: BthetaBtheta_App}\\
    B_{\theta} B_{\phi} &=& \tfrac{i}{2} \sum_{s,t} (h_{st}^{++}Y_{st}^{+2} - h_{st}^{--}Y_{st}^{-2}), \\
    B_{\phi} B_{\phi} &=& -\tfrac{1}{2}\,\sum_{s,t}  (h_{st}^{++}Y_{st}^{+2} + 2h_{st}^{+-}Y_{st}^{0} + h_{st}^{--}Y_{st}^{-2}).\label{eqn: BphiBphi_App}
\end{eqnarray}

\section{Representation of Splitting Data} \label{sec: a_coeff_Appendix}
It is standard practice in seismology and helioseismology to represent frequency splitting data by numbers called splitting coefficients (also called $a$ coefficients), which describe the decomposition of $\delta{}_n\omega{}_{\ell m} = {}_n\omega{}_{\ell m}-{}_n\omega{}_{\ell}$ in terms of some basis function over $m$ as follows
\begin{equation}
{}_n\omega{}_{\ell m} = {}_n\omega{}_{\ell} + \sum_{j=0}^{j_\text{max}} a^{n\ell}_j\, \cP_j(m),
\label{eq:a_def}
\end{equation}
where $a_{j}^{n\ell}$ are $a$ coefficients and $\cP_{j}(m)$ represents a $j_\text{max}+1$-dimensional orthogonal basis of polynomials on the discrete space of $m$'s which runs from $-\ell$ to $\ell$. In practice, $a$ coefficients are recorded for $j_\text{max}=36$ \citep[e.g.,][]{schou_data}. A recipe for obtaining these may be found in Appendix~A of \cite{schou_pol_94}. Harnessing the orthogonality of the basis polynomials, $\sum_{m=-\ell}^\ell \cP_{j}(m) \cP_{k}(m) = \delta_{jk}\, \sum_{m=-\ell}^\ell \left[\cP_{j}(m)\right]^2$, we write the $a$ coefficients as
\begin{equation}
a^{n\ell}_j = \sum_{m=-\ell}^\ell \delta{}_n\omega{}_{\ell m}\,\cP_j(m) \bigg/ \sum_{m=-\ell}^\ell \left[\cP_j(m)\right]^2.
\end{equation}



\section{Coupling of multiplets due to differential rotation} \label{sec: DR_mode_labelling}

One of the most precise measurements that helioseimology has offered is that of the differential rotation of the Sun  \citep{mjt03}. However, a primary premise in these inversions is that the isolated-multiplet assumption is valid. This is shown in \citet{lavely92} by calculating the coupling strength between two multiplets.
We performed an extensive calculation and also converge on the conclusion that the isolated-multiplet is a valid assumption for perturbations due to differential rotation. The rotation profile is dominantly axisymmetric with odd angular degree $s={1,3,5}$. The perturbing background velocity is
\begin{equation}
    \boldsymbol{u}_0(r) = \sum_{s=1,3,5} - w_s^0(r) \boldsymbol{\hat{r}} \times \boldsymbol{\bnabla_1} Y_s^0 (\theta,\phi),
\end{equation}
where $w_s^0(r)$ are the coefficients of the toroidal component of the axisymmetric vector spherical harmonics. The radial profile of solid body rotation is $w_1^0(r)$, which is 440~nHz at the surface, the equatorial surface rotation of the Sun.

\begin{figure}
\centering
    \includegraphics[totalheight=15cm]{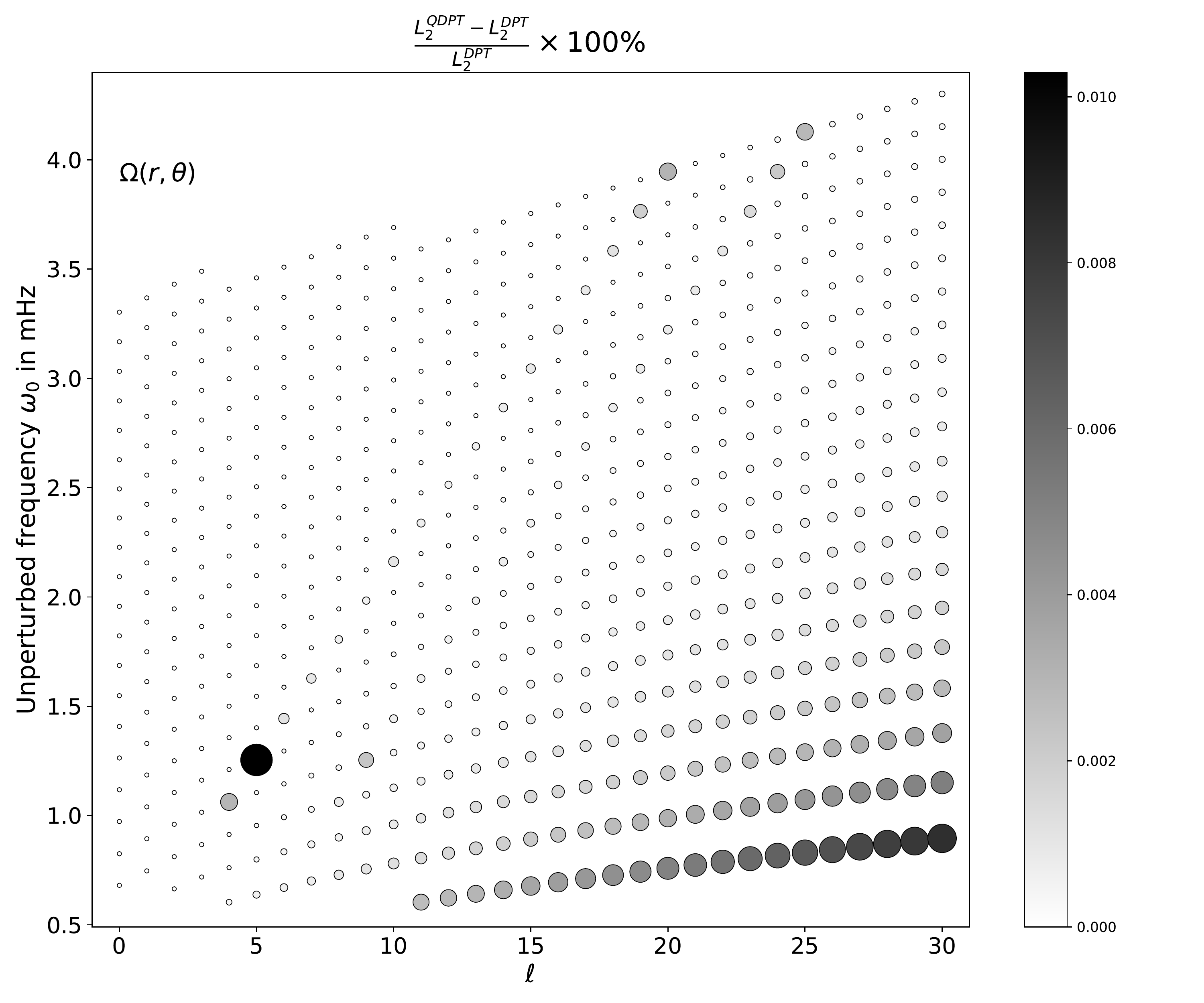}
    \caption{The relative offset of $L_2^\text{QDPT}$ as compared to that of $L_2^\text{DPT}$ (see Eqn~[\ref{eqn: l2_Q},\ref{eqn: l2_D}]) under the perturbation of an axisymmetric differential rotation $\Omega(r,\theta)$ as observed in the Sun. The gray-scale intensity and size of each `o' (representing a multiplet) increases with increasing departure of $\delta {}_{n}\omega{}_{\ell m}^\text{Q}$ from $\delta {}_{n}\omega{}_{\ell m}^\text{D}$. A larger and darker `o', implies stronger cross-coupling for that multiplet.}
    \label{fig:DR_coupling}
\end{figure}

Although differential rotation is a much stronger perturbation than Lorentz-stresses, there are no invisible modes (as opposed to what we found in Figure~\ref{fig: mode_labelling}). This is because the strong self-coupling of multiplets and therefore a dominantly diagonal supermatrix $\mathcal{Z}_{k'k}$. This results in normalized eigenvector corrections $c_{k'} \sim 1.0$ (refer Eqn~[\ref{eqn: general_Z_eigenproblem}]) only for $k' = k$, where $k=(n,\ell,m)$ and $k'=(n',\ell',m')$ are mode labels before and after perturbation respectively. Thus, even though $n,\ell$ do not stay good labels post perturbation, we can map the perturbed to the unperturbed modes by analyzing the eigenvectors. The azimuthal order $m$ remains a good label because of the axisymmetry of $\boldsymbol{u}_0(r)$.
Figure~\ref{fig:DR_coupling} has a maximum offset of 0.01\% when using cross-coupling as opposed self-coupling. Therefore, the validity of isolated multiplets holds good. Although Figure~\ref{fig:DR_coupling} shows degrees up to $\ell=30$, we calculated offsets up to $\ell=150$, and the maximum offset was 0.1\%, located in the fundamental branch $n=0$. 

\bibliography{ms}{}
\bibliographystyle{apa}






\end{document}